\documentclass[a4paper,12pt,onecolumn]{IEEEtran}

\usepackage{graphicx}
\usepackage{subfigure}
\usepackage{amssymb,amsmath}
\usepackage{bm,bbm}
\usepackage{booktabs}
\usepackage{subfigure}
\usepackage{microtype}
\usepackage{color}
\usepackage{cite}
\usepackage{balance}

\newtheorem{proposition}{Proposition}
\begin{document}
\title{When Data do not Bring Information: A Case Study in Markov Random Fields Estimation}

\author{Javier~Gimenez,
        Alejandro~C.~Frery,~\IEEEmembership{Senior~Member,~IEEE}
        and~Ana~Georgina~Flesia
\thanks{J.~Gimenez is with Conicet and INAUT --\textit{ Instituto de Autom\'atica, Facultad de Ingenier\'ia}, UNSJ -
\textit{Universidad Nacional de San Juan}, Av. Lib. San Mart\'in (Oeste) 1109, J5400ARL, San Juan, Argentina.
jgimenez@inaut.unsj.edu.ar}
\thanks{A.~C.~Frery is with LaCCAN -- \textit{Laborat\'orio de Computa\c c\~ao Cient\'ifica e An\'alise Num\'erica}, Ufal - \textit{Universidade Federal de Alagoas}, Macei\'o, Brazil. acfrery@gmail.com.}
\thanks{A.~G.~Flesia is with Conicet and FaMAF - \textit{Facultad de Matem\'atica, Astronom\'ia y F\'isica}, UNC - \textit{Universidad Nacional de C\'ordoba}, Av. Medina Allende, s/n, X5000HUA, C\'ordoba, Argentina.
flesia@mate.uncor.edu}}

\markboth{To appear at IEEE JOURNAL OF SELECTED TOPICS IN APPLIED EARTH OBSERVATIONS AND REMOTE SENSING,}%
{Shell \MakeLowercase{\textit{et al.}}: When Data do not Bring Information}

\maketitle
\newtheorem{prop}{Proposition}
\newenvironment{DEM}{\bf Dem. \rm}{$\hfill \Box$}

\newcommand{\Keywords}[1]{\par\noindent
{\small{\em Index Terms\/}: #1}}

\begin{abstract}
The Potts model is frequently used to describe the behavior of image classes, since it allows to incorporate contextual information linking neighboring pixels in a simple way.
Its isotropic version has only one real parameter $\beta$, known as smoothness parameter or inverse temperature, which regulates the classes map homogeneity.
The classes are unavailable, and estimating them is central in important image processing procedures as, for instance, image classification.
Methods for estimating the classes which stem from a Bayesian approach under the Potts model require to adequately specify a value for $\beta$.
The estimation of such parameter can be efficiently made solving the Pseudo Maximum likelihood (PML) equations in two different schemes, using the prior or the posterior model.
Having only radiometric data available, the first scheme needs the computation of an initial segmentation,  while the second uses both the segmentation and the radiometric data to make the estimation.
In this paper, we compare these two PML estimators by computing the mean square error (MSE), bias, and sensitivity to deviations from the hypothesis of the model.
We conclude that the use of extra data does not improve the accuracy of the PML, moreover, under gross deviations from the model, this extra information introduces unpredictable distortions and bias.
\end{abstract}

\section{Introduction}

Geman and Geman~\cite{geman1984} consolidated the use of Gibbs laws as prior evidence in the processing and analysis of images.
Such distributions are able to capture the spatial structure of the visual information in a tractable manner.
Among them, the Potts model has become a commonplace for describing classes.
In its simplest isotropic version, the amount of spatial association is controlled by the smoothness parameter $\beta$ which is a real value also known as smoothness parameter or inverse temperature.
Within the Bayesian framework, assuming the Potts model as the prior distribution for the classes, the posterior distribution of the class map given the radiometric data is also a Potts model, in which the likelihood of the observed data appears as an external field.
Moreover, the contextual information is also described by the same $\beta$, albeit the particular form of the law is not the same~\cite{bustos1992}.

Many estimators of the true (unobserved) map of classes given the observations can be proposed in this context.
Among them, MAP (Maximum A Posteriori), MPM (Maximum Posterior Marginals) and ICM (Iterated Conditional Modes) stem as natural procedures.
Computing any of the two first is an NP problem, so approximate procedures have been proposed, e.g. Simulated Annealing~\cite{geman1984} and incomplete estimators~\cite{Ferrari1995}, whereas the ICM~\cite{besag1986} is an attractive estimator which leads to an iterative classification procedure.
All these techniques require to adequately specify a value for $\beta$.
In the literature,  for the ICM estimator, the specification of $\beta$ has been diverse: fixing $\beta$ by trial-and-error~\cite{Arbia1999,Jackson2002}, estimating $\beta$ once and keeping this value until convergence~\cite{Descombes1999,Melgani2003,Tso1999}, or updating the value for each iteration~\cite{Frery2007,Frery2009}.

Classical statistical estimators of the smoothness parameter require computing or estimating the normalizing constant $Z_\beta$ of the Potts model, which is generally quite difficult.
Exact recursive expressions have been proposed to compute it analytically \cite{McGrory2009}.
However, to our knowledge, these recursive methods have only been successfully applied to small problems.
General Monte Carlo Markov Chain methods can not be applied to estimate $\beta$, but some specific MCMC algorithms have been designed in \cite{liu2000,Risser2011}, among others.

Another popular estimation method applies the EM (Expectation Maximization) algorithm to approximate maximum likelihood estimators.
This procedure iterates between the computation of the expectation of the joint log-likelihood of the data and the labels, given $\beta$ (step E) and the update of $\beta$ as the argument that maximizes the expectation (step M). Several EM optimizations may be found on \cite{Ibanez2003} and references therein.

In all these previous methods, the partition function $Z_{\beta}$ or an estimate of it is needed, while other methods work independently of $Z_{\beta}$.
In \cite{Ali2008}, the first three terms of the Taylor series around $\beta=0$  of the log-likelihood  are found,  and $\beta$ is computed as the argument that maximizes that formula.
In \cite{Pereyra2013}, the estimation of $\beta$ was included within an MCMC method using an approximative Bayesian computation likelihood-free Metropolis-Hastings algorithm, in which $Z_{\beta}$ was replaced by a simulation-rejection scheme.

All these methods are inaccurate or computationally expensive, with exception of the PML estimators \cite{Besag1975}.
These last estimators circumvent the use of $Z_{\beta}$, replacing the likelihood function by a product of conditional densities, which may come from the prior or posterior model.

The pseudo likelihood estimator  based on the prior model is a classical estimator which has been often applied in contextual image segmentation methods, see \cite{Frery2007,Levada2008,Flesia2013a,Flesia2013b} for details. In \cite{Gimenez2013}, a new PML estimator was introduced, which is based on the posterior model. The estimation in both methods is performed over an observed segmentation, since it estimates the smoothness of the class configuration.
In principle, the advantage of using the posterior distribution is that the radiometric data is also included in the estimation process.

Gimenez et al.~\cite{Gimenez2013} showed that adding radiometric data into the PML equation produces highly accurate estimates, i.e. with negligible Mean Square Error (MSE).
Also, working with real Landsat data, estimation using the posterior model seemed to improve the ICM segmentation output.
Nevertheless, when studying more closely the behavior of the estimate under a contaminated model, i.e. when the smoothness of the initial segmentation does not agree with a Potts model,  great instability was discovered in outputs of the estimator based on the posterior distribution that where not shared by the classical estimate based only in the prior distribution.

To the best of our knowledge, there is no discussion in the literature about the stability of PML estimators of the Pott's  smoothness parameter under data contamination.
This finding gives more relevance to the version of the ICM algorithm discussed  in~\cite{Frery2009}, which estimates the smoothness parameter each time the configuration is updated, since the initial class configuration (obtained from noisy data) may introduce great bias in the smoothness estimate, resulting in severe underestimation of the influence of the context on the final result.

This paper is organized as follows.
In Section~\ref{Sec:Definitions} a general review of  the Potts model, general Pseudolikelihood estimation and simulation techniques for such model is made.
In Section~\ref{Sec:Estimators}, the two PML estimators are compared using data simulated under the Potts model, analyzing Mean Square Error, Bias and Variance.
In Section~\ref{Sec:Sensitivity}, the influence of the smoothness of the initial segmentation is studied, considering contextual ICM and Maximum likelihood classification with Gaussian observation in each class.
Conclusions are drawn in Section~\ref{Sec:Conclusions}.

\section{Definitions}\label{Sec:Definitions}

\subsection{Model}

Without loss of generality, a finite image is a function defined on a grid $S$ of $n$ lines and $m$ columns.
A Bayesian model stipulates that at each position $s\in S$ there is an element from the set of possible classes $\mathcal{L}=\{\ell_1,\ldots,\ell_L\}$, $L\geq2$.
The random field which describes all the classes is denoted by $\bm X=(X_s)_{s\in S}$, and its distribution is called the prior distribution.
Assuming that the observations, given the classes, are independent random variables, the observed image can be described in conditional terms by a probability law $p(\cdot\mid x_s)$ which depends only on the observed class at $s$.
The conditional laws $\{p(\cdot\mid\ell):\ell\in\mathcal{L}\}$ are the distribution of the data $\bm I$ given the classes $\bm X$.
The Bayesian classification problem consists of estimating $\bm X$ provided $\bm I$.

The Potts model is one of the most widely used prior distributions in Bayesian image analysis.
The basic idea is that the distribution of $X_s$ conditioned on the rest of the field only depends of the class configuration on a (usually small with respect to $n$ and $m$) set of neighbors $\partial_s\subset S$.
All neighbors form the neighborhood of the field, which has the following properties:
(i)~$s\notin\partial_s$, (ii)~$s\in\partial_t\iff t\in\partial_s$, and (iii)~$S=\cup_{s\in S}\partial_s$.

In the isotropic and without external field version of this model, the probability of observing class $\ell\in\mathcal{L}$ in any coordinate $s$ given the classes in its neighborhood $x_{\partial_s}$ is given by
\begin{equation}
f_{X_s\mid X_{\partial_s},\beta}(\ell\mid x_{\partial_s})\propto\exp\{\beta U_s(\ell)\},\label{eq:ConditionalProbabilities}
\end{equation}
where $U_s(\ell)$ is the number of neighbors of $s\in S$ with label $\ell\in\mathcal{L}$.
These conditional probabilities uniquely specify the joint distribution of $\bm X$,
\begin{equation}
f_{\bm X,\beta}(\bm x)\propto\exp\{\beta U(\bm x)\},
\label{prior}
\end{equation}
where $U(\bm x)$ is the number of pairs of neighboring pixels with the same label in the class map $\bm x$.

We are interested in the case $\beta>0$ which promotes spatial smoothness, a desirable property for the prior distribution in classification procedures, but the forthcoming discussion is analogous for the case $\beta<0$.

Assuming that the observations, given the classes, are independent random variables, applying the Bayes rule one obtains the distribution of the classes $\bm X$ given the observations $\bm I$:
\begin{equation}
f_{\bm X\mid \bm I,\beta}(\bm x\mid \bm I)\propto\exp\bigg\{\sum_{s\in S}\ln p(I_s\mid x_s)+\beta U(\bm x)\bigg\}.
\label{posteriori}
\end{equation}
This is the Potts model subjected to the external field $(\ln p(I_s\mid x_s))_{s\in S}$.

\subsection{Inference}

The joint distribution of the Potts model, either for the classes only or for the posterior distribution, involves an unknown normalization constant, $Z_{\beta}$, termed ``partition function'' in the literature.
Since this constant depends on the smoothness parameter, iterative algorithms for computing a maximum likelihood estimator of $\beta$ require evaluating this function a number of times, which is unfeasible in practical situations. Pseudolikelihood estimators are an interesting alternative to solve this problem.
Instead of finding the parameter which maximizes the joint distribution, they are defined as the argument which maximizes the product of conditional distributions.

Given $\bm x$, an observation of the model characterized by (\ref{prior}), a classical proposal $\widehat{\beta}_{\text{prior}}$ consists in solving the maximum pseudolikelihood equation
$$\widehat{\beta}_{\mbox{\scriptsize prior}}=\arg\max_{\beta}\prod_{s\in S}f_{X_s\mid X_{\partial_s}}(x_s\mid x_{\partial_s}).$$
After a series of algebraical steps, this is given by
\begin{equation}
f_{\mbox{\scriptsize prior}}(\widehat{\beta}_{\mbox{\scriptsize prior}})=0,
\label{ecuacion_pri}
\end{equation}
where
\begin{equation}
f_{\mbox{\scriptsize prior}}(\beta)=\sum_{s\in S}U_s(x_s)-\sum_{s\in S}\frac{\sum_{\ell\in\mathcal{L}}U_s(\ell)\exp\{\beta U_s(\ell)\}}{\sum_{\ell\in\mathcal{L}}\exp\{\beta U_s(\ell)\}}.
\label{f_prior}
\end{equation}
If $\partial_s$ consists of the eight closest neighbors (discarding sites by the edges and corners of $S$), equation~(\ref{ecuacion_pri}) reduces to the nonlinear equation
in $\beta$ with only 23 terms given in (\ref{eq:PseudoPottsPuro}), where the coefficients $K_i$ count the number of patches with certain configurations; see \cite{Levada2009} for details.
\begin{figure*}[hbt]
\hrulefill
{\small
\begin{align}
0=&\sum_{s\in S}U_s(x_s)-\frac{8e^{8\widehat{\beta}}}{e^{8\widehat{\beta}}+L-1}K_1
-\frac{7e^{7\widehat{\beta}}+e^{\widehat{\beta}}}{e^{7\widehat{\beta}}+e^{\widehat{\beta}}+L-2}K_2
-\frac{6e^{6\widehat{\beta}}+2e^{2\widehat{\beta}}}{e^{6\widehat{\beta}}+e^{2\widehat{\beta}}+L-2}K_3
-\frac{6e^{6\widehat{\beta}}+2e^{\widehat{\beta}}}{e^{6\widehat{\beta}}+2e^{\widehat{\beta}}+L-3}K_4 \nonumber\\
&{}-\frac{5e^{5\widehat{\beta}}+3e^{3\widehat{\beta}}}{e^{5\widehat{\beta}}+e^{3\widehat{\beta}}+L-2}K_5
-\frac{5e^{5\widehat{\beta}}+2e^{2\widehat{\beta}}+e^{\widehat{\beta}}}{e^{5\widehat{\beta}}+e^{2\widehat{\beta}}+e^{\widehat{\beta}}+L-3}K_6
-\frac{5e^{5\widehat{\beta}}+3e^{\widehat{\beta}}}{e^{5\widehat{\beta}}+3e^{\widehat{\beta}}+L-4}K_7
-\frac{8e^{4\widehat{\beta}}}{2e^{4\widehat{\beta}}+L-2}K_8\nonumber\\
&{}-\frac{4e^{4\widehat{\beta}}+3e^{3\widehat{\beta}}+e^{\widehat{\beta}}}{e^{4\widehat{\beta}}+e^{3\widehat{\beta}}+e^{\widehat{\beta}}+L-3}K_9
-\frac{4e^{4\widehat{\beta}}+4e^{2\widehat{\beta}}}{e^{4\widehat{\beta}}+2e^{2\widehat{\beta}}+L-3}K_{10}
-\frac{4e^{4\widehat{\beta}}+2e^{2\widehat{\beta}}+2e^{\widehat{\beta}}}{e^{4\widehat{\beta}}+e^{2\widehat{\beta}}+2e^{\widehat{\beta}}+L-4}K_{11}\nonumber\\
&{}-\frac{4e^{4\widehat{\beta}}+4e^{\widehat{\beta}}}{e^{4\widehat{\beta}}+4e^{\widehat{\beta}}+L-5}K_{12}
-\frac{6e^{3\widehat{\beta}}+2e^{2\widehat{\beta}}}{2e^{3\widehat{\beta}}+e^{2\widehat{\beta}}+L-3}K_{13}
-\frac{6e^{3\widehat{\beta}}+2e^{\widehat{\beta}}}{2e^{3\widehat{\beta}}+2e^{\widehat{\beta}}+L-4}K_{14} \nonumber\\
&{}-\frac{3e^{3\widehat{\beta}}+4e^{2\widehat{\beta}}+e^{\widehat{\beta}}}{e^{3\widehat{\beta}}+2e^{2\widehat{\beta}}+e^{\widehat{\beta}}+L-4}K_{15}
-\frac{3e^{3\widehat{\beta}}+2e^{2\widehat{\beta}}+3e^{\widehat{\beta}}}{e^{3\widehat{\beta}}+e^{2\widehat{\beta}}+3e^{\widehat{\beta}}+L-5}K_{16}
-\frac{3e^{3\widehat{\beta}}+5e^{\widehat{\beta}}}{e^{3\widehat{\beta}}+5e^{\widehat{\beta}}+L-6}K_{17}\nonumber\\
&{}-\frac{8e^{2\widehat{\beta}}}{4e^{2\widehat{\beta}}+L-4}K_{18}
-\frac{6e^{2\widehat{\beta}}+2e^{\widehat{\beta}}}{3e^{2\widehat{\beta}}+2e^{\widehat{\beta}}+L-5}K_{19}
-\frac{4e^{2\widehat{\beta}}+4e^{\widehat{\beta}}}{2e^{2\widehat{\beta}}+4e^{\widehat{\beta}}+L-6}K_{20}
-\frac{2e^{2\widehat{\beta}}+6e^{\widehat{\beta}}}{e^{2\widehat{\beta}}+6e^{\widehat{\beta}}+L-7}K_{21}\nonumber\\
&{}-\frac{8e^{\widehat{\beta}}}{8e^{\widehat{\beta}}+L-8}K_{22}.
\label{eq:PseudoPottsPuro}
\end{align}}
\hrulefill	
\end{figure*}

Equation (\ref{ecuacion_pri}) involves an observed map of classes $\bm x$, that the model should follow in order to allow adequate parameter estimation.
This is, the map of classes should be a realization of a Potts model.
Nevertheless, in practice, only the radiometric image data are available, thus  the map of classes must be firstly estimated from the image by a classification method, and then parameter estimation can be pursued.
As observed in~\cite{Frery2007}, the initial map of classes $\bm x$ is usually obtained by maximum likelihood classification from the image data, assuming no spatial structure for the classes, i.e., $\beta=0$.
The radiometric data has, therefore, influence on the estimation process, and such information should not be discarded before checking the degree of influence in the accuracy of the estimation.
In \cite{Gimenez2013}, following this approach, a new estimator $\widehat{\beta}_{\text{post}}$  was proposed, incorporating the observed radiometric data into the estimation itself by considering the posterior conditional distributions.
Thus, the estimator was defined as the solution of the following equation:
$$\widehat{\beta}_{\mbox{\scriptsize post}}=\arg\max_{\beta}\prod_{s\in S}f_{X_s\mid I_s,X_{\partial_s}}(x_s\mid I_s,x_{\partial_s}).$$
Such equation can be transformed into
\begin{equation}
f_{\mbox{\scriptsize post}}(\widehat{\beta}_{\mbox{\scriptsize post}})=0,
\label{ecuacion_post}
\end{equation}
where
\begin{align}
f_{\mbox{\scriptsize post}}(\beta)=&\displaystyle\sum_{s\in S}U_s(x_s)-\nonumber\\
&-\sum_{s\in S}\frac{\sum_{\ell\in\mathcal{L}}U_s(\ell)p(I_s\mid \ell)\exp\{\beta U_s(\ell)\}}{\sum_{\ell\in\mathcal{L}}p(I_s\mid \ell)\exp\{\beta U_s(\ell)\}}.
\label{f_post}
\end{align}

If $\bm x$ is a map for which there are two pixels $s,t\in S$ such that
\begin{equation}
U_s(x_s)>\min_{\ell\in\mathcal{L}}U_s(\ell)\hspace{0.4cm}\mbox{and}\hspace{0.4cm}U_t(x_t)<\max_{\ell\in\mathcal{L}}U_t(\ell),
\label{condicion_ex_unic}
\end{equation}
then equations (\ref{ecuacion_pri}) and (\ref{ecuacion_post}) have a unique solution, since the functions (\ref{f_prior}) and (\ref{f_post}) are strictly decreasing (See Proposition~1), continuous and they verify
\begin{align*}
\lim_{\beta\rightarrow-\infty}f_{\mbox{\scriptsize prior}}(\beta)=\lim_{\beta\rightarrow-\infty}f_{\mbox{\scriptsize post}}(\beta)=\\
=\sum_{s\in S}(U_s(x_s)-\min_{\ell\in\mathcal{L}}U_s(\ell))>0,
\end{align*}
and
\begin{align*}
\lim_{\beta\rightarrow\infty}f_{\mbox{\scriptsize prior}}(\beta)=\lim_{\beta\rightarrow\infty}f_{\mbox{\scriptsize post}}(\beta)=\\
=\sum_{s\in S}(U_s(x_s)-\max_{\ell\in\mathcal{L}}U_s(\ell))<0.
\end{align*}

\begin{figure}[hbt]
\centering
\subfigure{
  \includegraphics[scale = 0.24]{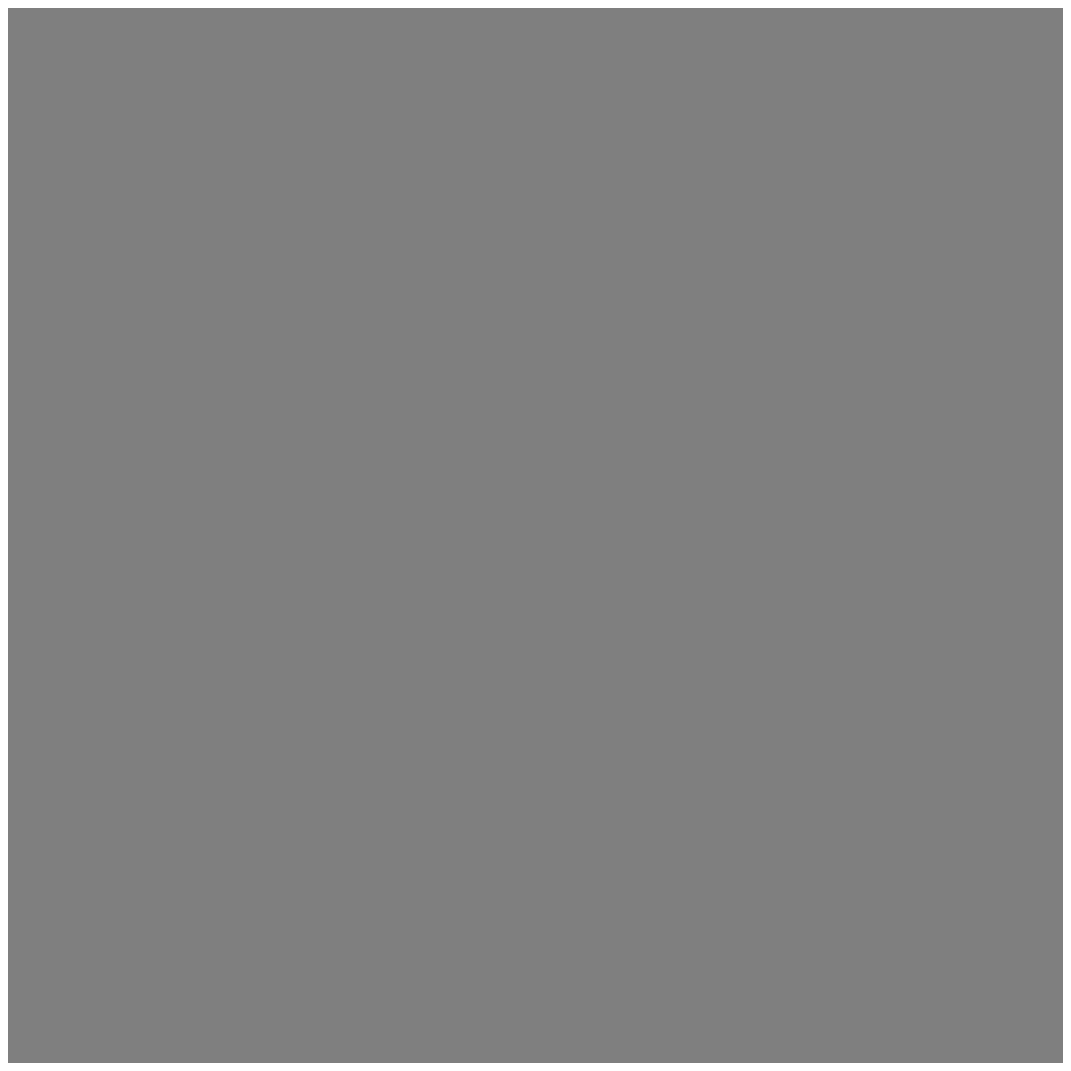}
}
\subfigure{
  \includegraphics[scale = 0.24]{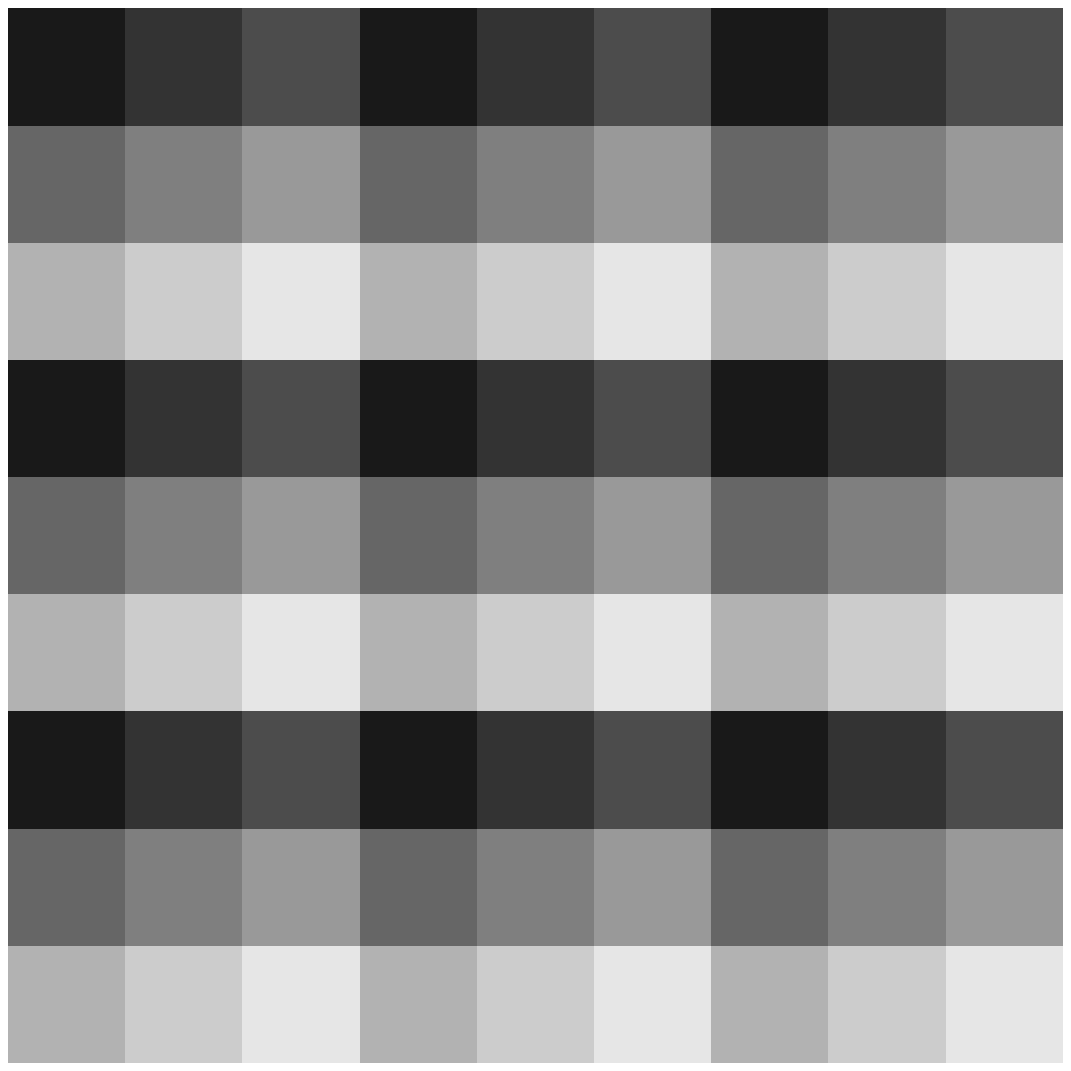}
}
\subfigure{
  \includegraphics[scale = 0.24]{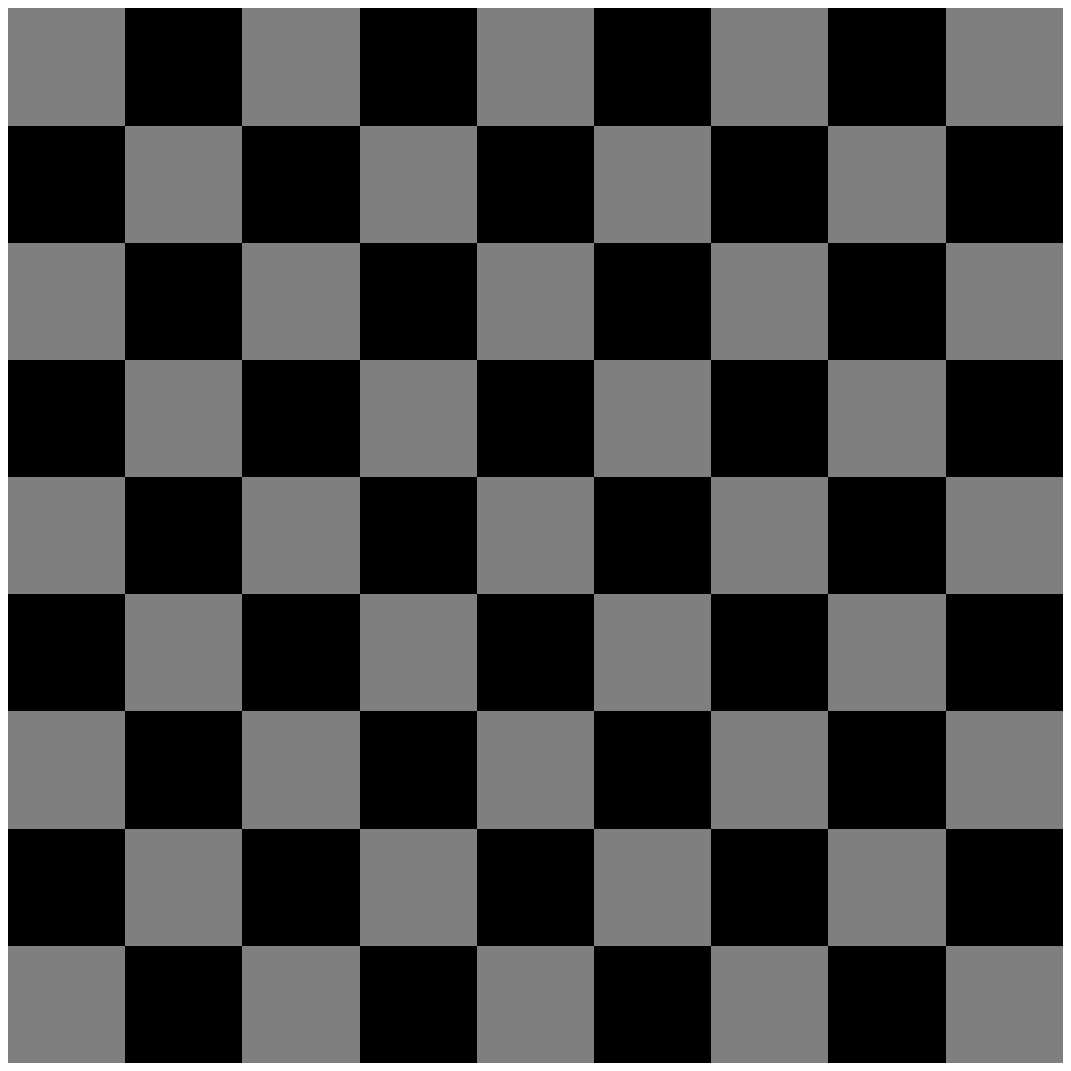}
}
\caption{Class maps for which condition (\ref{condicion_ex_unic}) do not hold for second order neighborhoods.}
\label{casos_raros}
\end{figure}

The conditions given in  (\ref{condicion_ex_unic}) are usually verified except in rare cases like the $9\times9$ maps shown in Fig.~\ref{casos_raros} for second order neighborhoods.

Thus we have presented two PML estimators of $\beta$, one that needs only the map of classes, and other that uses the map and the observed image.

In this paper we will show that, for the Potts model, the intuitive claim that more information implies better estimation does not hold.
The use of the extra information contained in the posterior model~(\ref{posteriori}), does not improve the estimations under the model given in equation~(\ref{prior}), moreover, it makes estimation more sensitive to deviations from the Potts model.
This means that the observed (or estimated) map of classes is sufficient to obtain accurate smoothness parameter estimations under the true model, and it maintains the accuracy under common deviations of the model, which are usual in practice.

\subsection{Simulation}

We study the behavior of the two PML estimators under the pure model (class maps and radiometric data are simulated), and under contaminated model (estimated classifications instead of the simulated class maps).

There are many well known algorithms for simulating realizations of the second order Potts model.
We have implemented our version of the  Swendsen-Wang algorithm \cite{SW1987} on Matlab; the computational details are in the Appendix.
We generated $100$ realizations of the Potts model of size $128\times128$ for each combination of parameters $\beta$ and number of classes $L$ in the sets  $\beta\in\{0.1$; $0.2$; $0.3$; $0.4$; $0.45$; $0.5$; $0.6$; $0.7$; $0.8$; $0.9$; $1\}$, $L\in\{2,3,4\}$.
For each simulated class map, Gaussian radiometric data was also simulated with the same variance per class, but  means separated by $k=1$, $2$, $3$ and $4$ standard deviations.
Each Gaussian image  was classified with (Gaussian) maximum likelihood with the true emission parameters, and this (non contextual) classification was then used to initialize the Iterated Conditional Modes algorithm with $\beta$ set as the parameter that generated the simulated map of classes.

\begin{figure}[h!]
\centering
\subfigure[Potts Model realization.\label{fig_potts}]{
  \includegraphics[scale = 0.45]{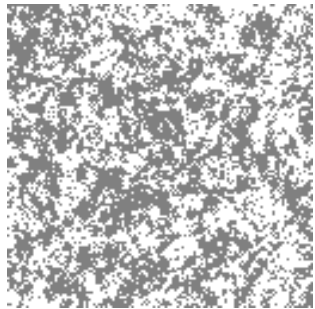}
}
\subfigure[Gaussian data.\label{fig_datos}]{
  \includegraphics[scale = 0.45]{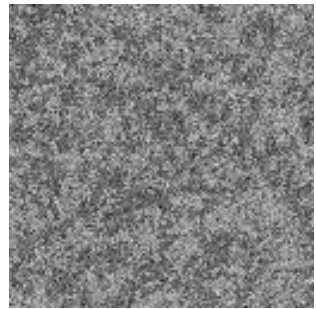}
}

\subfigure[Histogram of (b) for each class according to (a) and the corresponding to the full image (b).\label{fig_hist}]{
  \includegraphics[scale = 0.36]{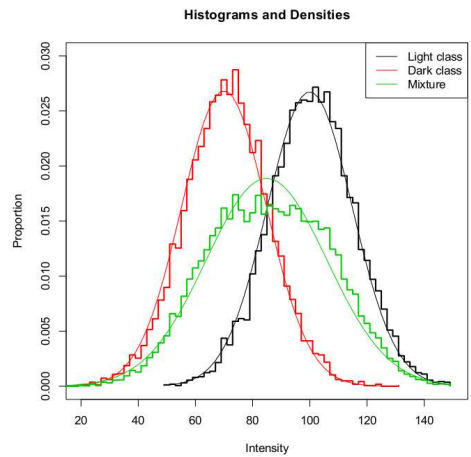}
}

\subfigure[ML of (b).\label{fig_ML}]{
  \includegraphics[scale = 0.45]{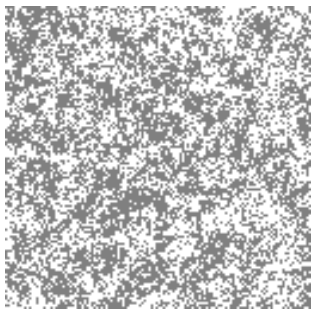}
}
\subfigure[ICM with $\beta=0.36$ fixed and initial clasification (d).\label{fig_ICM}]{
  \includegraphics[scale = 0.45]{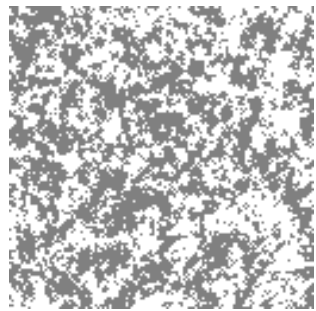}
}
\caption{Simulated data base example for $L=2$; $\beta=0.3$; and $k=2$.}
\label{simulations}
\end{figure}

Fig.~\ref{simulations} shows an example with $L=2$, $\beta=0.3$ and $k=2$; Fig.~\ref{fig_potts} a realization of the  Potts model, while Fig.~\ref{fig_datos} presents the Gaussian observed data with standard deviation $15$ and means $70$ and $100$ respectively (separated by $k=2$ standard deviations).
Fig.~\ref{fig_hist} shows the histogram of the emission model of each class and of the mixture of classes.
Fig~\ref{fig_ML} shows the Gaussian ML classification of the observed data; and Fig.~\ref{fig_ICM} presents their ICM segmentation using Fig.~\ref{fig_ML} and $\beta=0.3$ as initial map.

The PML estimators are the roots of the functions $f_{\mbox{\scriptsize prior}}$ and $f_{\mbox{\scriptsize post}}$.
We will use simulated data to plot these functions since the analysis of such curves is important to explain the numerical instabilities that are introduced by common, simple deviations from the model.
In Fig.~\ref{fig_fprior_fpost} we show an example of the curves corresponding to $L=3$; $\beta=0.4$; and $k=1$.
The $\beta$ axis and the true value of the parameter corresponding to the model are respectively marked by a vertical and horizontal solid line.
Thus, the estimation is good if the curve passes where the vertical and horizontal lines intersect.
In the example of Fig.~\ref{fig_fprior_fpost}, both estimators are accurate.
We should note that the plot of the curves $f_{\mbox{\scriptsize prior}}$ and  $f_{\mbox{\scriptsize post}}$ are very smooth, albeit their complex algebraic expressions.

\section{Statistical Accuracy: MSE, bias and variance}\label{Sec:Estimators}

\begin{figure}[hbt]
\centering
\subfigure[$f_{\mbox{\scriptsize prior}}$ and $f_{\mbox{\scriptsize post}}$\label{fig_fprior_fpost}]{
  \includegraphics[scale = 0.23]{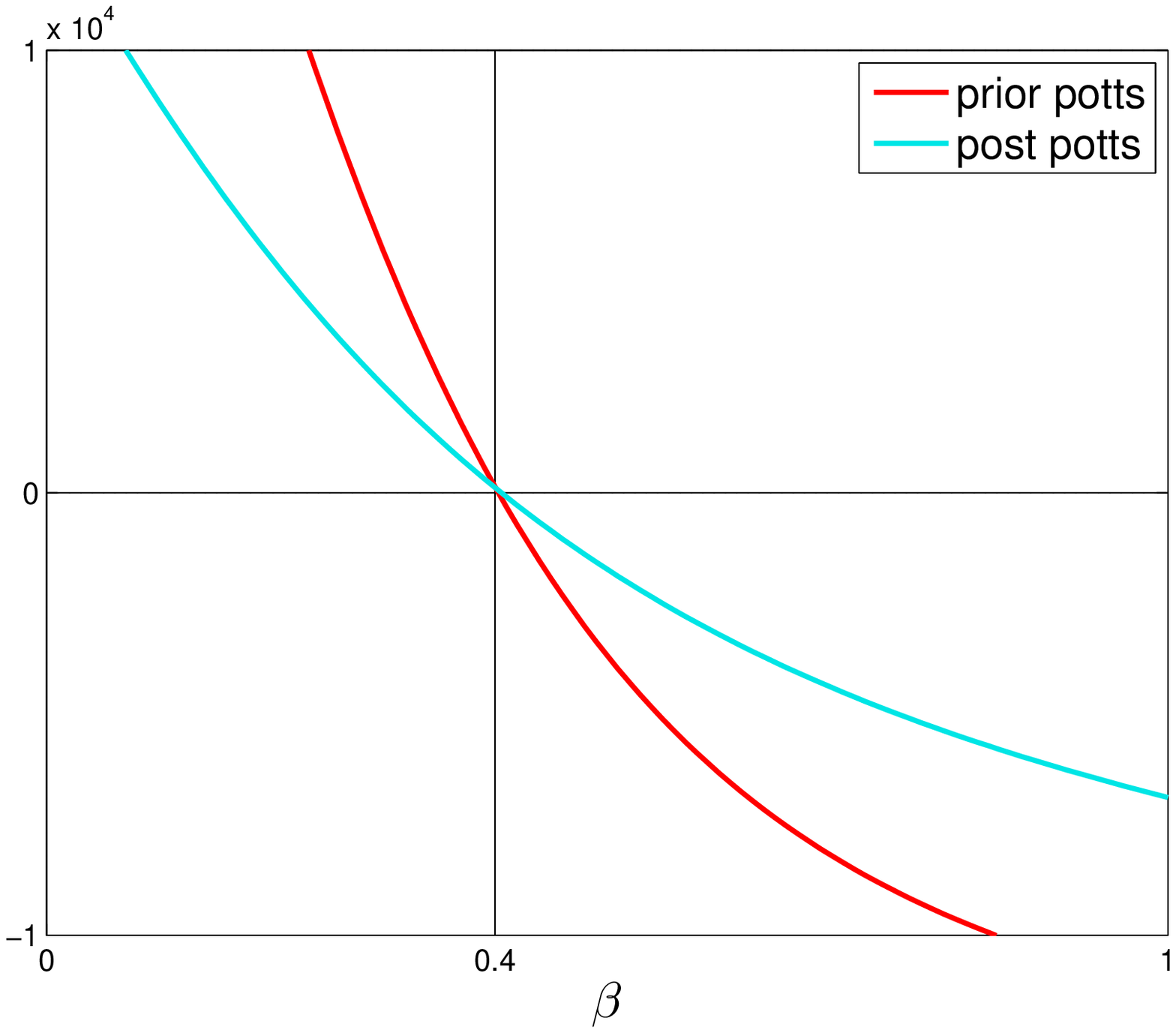}
}
\subfigure[Curve Bundles for $f_{\mbox{\scriptsize prior}}$ and $f_{\mbox{\scriptsize post}}$\label{fig_haces_fprior_fpost}]{
  \includegraphics[scale = 0.23]{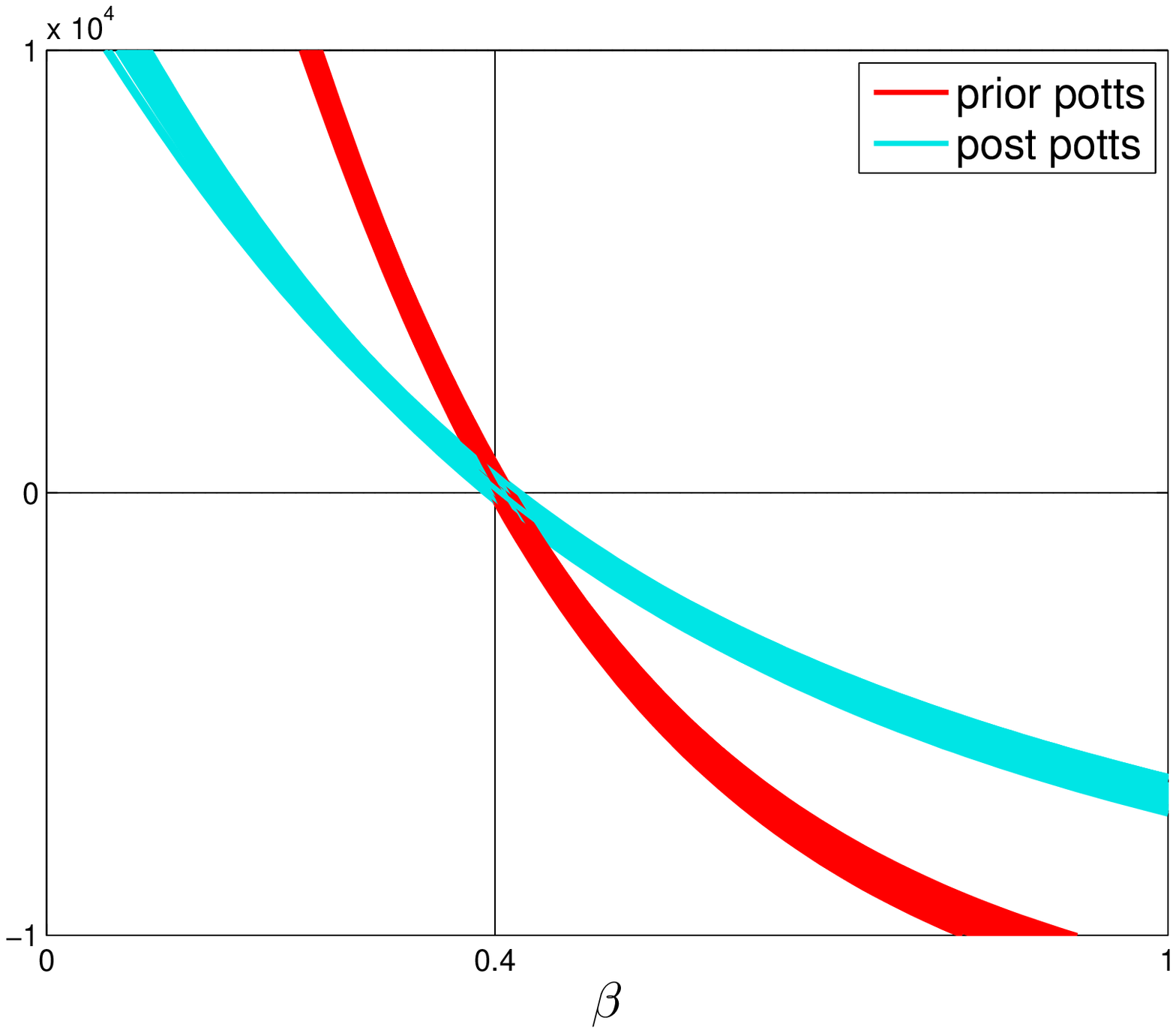}
}
\caption{Functions whose roots are the PML estimators ($f_{\mbox{\scriptsize prior}}$ and $f_{\mbox{\scriptsize post}}$), along with their bundles for the case $L=3$; $\beta=0.4$ and $k=2$.}
 \label{grafico_funciones}
\end{figure}

We analyze the Mean Square Error (MSE), the bias and the variance of the estimators, computed on the simulated data under the true Potts model described in the previous section.
This information is presented in Table~\ref{Tabla1}.
Both accuracy and precision are good in both estimators, since there is no noticeable bias, nor large variance.
We also analyze these results plotting the one hundred curves $f_{\mbox{\scriptsize prior}}$ and $f_{\mbox{\scriptsize post}}$, for each choice of parameters.
Fig.~\ref{fig_haces_fprior_fpost} presents such bundles of curves for $L=3$, $\beta=0.4$, and $k=2$.
We conclude that under the model, both estimators are consistent and statistically indistinguishable.

\begin{table}[htb]
\caption{MSE, standard deviation and mean of estimators computed over $100$ realizations of a Potts model for several values of $\beta$, Gaussian data with $\sigma=15$ and $k=1$.}
\label{Tabla1}
\centering
\scriptsize
\begin{tabular}{@{}cccccccccc@{}} \toprule
&\multicolumn{9}{c}{$L=2$}\\ \hline
&&\multicolumn{2}{c}{$\sqrt{MSE}$} &&\multicolumn{2}{c}{Mean}&&\multicolumn{2}{c}{STD}\\
\cmidrule{3-4}\cmidrule{6-7}\cmidrule{9-10}
$\beta$ &&$\widehat{\beta}\mbox{ prior}$&$\widehat{\beta}\mbox{ post}$&&$\widehat{\beta}\mbox{ prior}$&$\widehat{\beta}\mbox{ post}$&&$\widehat{\beta}\mbox{ prior}$&$\widehat{\beta}\mbox{ post}$\\\hline
0.1   &&      0.006  &  0.007 && 0.101   &  0.102 && 0.006  &  0.006\\
0.2   &&      0.006  &  0.007 && 0.203   &  0.203 && 0.005  &  0.006\\
0.3   &&      0.008  &  0.007 && 0.306   &  0.306 && 0.005  &  0.005\\
0.4   &&      0.005  &  0.005 && 0.403   &  0.403 && 0.004  &  0.004\\
0.45  &&      0.005  &  0.006 && 0.452   &  0.452 && 0.005  &  0.005\\
0.5   &&      0.006  &  0.006 && 0.499   &  0.499 && 0.006  &  0.006\\
0.6   &&      0.040  &  0.039 && 0.609   &  0.612 && 0.038  &  0.037\\
0.7   &&      0.019  &  0.021 && 0.685   &  0.683 && 0.012  &  0.013\\
0.8   &&      0.034  &  0.036 && 0.771   &  0.769 && 0.018  &  0.019\\
0.9   &&      0.053  &  0.051 && 0.852   &  0.854 && 0.024  &  0.023\\
1     &&      0.082  &  0.082 && 0.923   &  0.924 && 0.032  &  0.033\\
 \toprule
&\multicolumn{9}{c}{$L=3$}\\ \hline
&&\multicolumn{2}{c}{$\sqrt{MSE}$} &&\multicolumn{2}{c}{Mean}&&\multicolumn{2}{c}{STD}\\
\cmidrule{3-4}\cmidrule{6-7}\cmidrule{9-10}
$\beta$ &&$\widehat{\beta}\mbox{ prior}$&$\widehat{\beta}\mbox{ post}$&&$\widehat{\beta}\mbox{ prior}$&$\widehat{\beta}\mbox{ post}$&&$\widehat{\beta}\mbox{ prior}$&$\widehat{\beta}\mbox{ post}$\\\hline
0.1    &&    0.007  &  0.008 && 0.100  & 0.100 && 0.007  &  0.008\\
0.2    &&    0.007  &  0.008 && 0.202  & 0.203 && 0.007  &  0.007\\
0.3    &&    0.006  &  0.007 && 0.304  & 0.304 && 0.004  &  0.005\\
0.4    &&    0.007  &  0.008 && 0.406  & 0.406 && 0.004  &  0.005\\
0.45   &&    0.004  &  0.004 && 0.453  & 0.452 && 0.002  &  0.003\\
0.5    &&    0.004  &  0.005 && 0.501  & 0.501 && 0.004  &  0.004\\
0.6   &&      0.065  &  0.068 && 0.648   &  0.652 && 0.043  &  0.043\\
0.7   &&      0.019  &  0.018 && 0.692   &  0.695 && 0.018  &  0.017\\
0.8   &&      0.028  &  0.029 && 0.774   &  0.774 && 0.013  &  0.014\\
0.9   &&      0.047  &  0.044 && 0.855   &  0.859 && 0.017  &  0.018\\
1     &&      0.074  &  0.071 && 0.929   &  0.932 && 0.022  &  0.022\\
\toprule
&\multicolumn{9}{c}{$L=4$}\\ \hline
&&\multicolumn{2}{c}{$\sqrt{MSE}$} &&\multicolumn{2}{c}{Mean}&&\multicolumn{2}{c}{STD}\\
\cmidrule{3-4}\cmidrule{6-7}\cmidrule{9-10}
$\beta$ &&$\widehat{\beta}\mbox{ prior}$&$\widehat{\beta}\mbox{ post}$&&$\widehat{\beta}\mbox{ prior}$&$\widehat{\beta}\mbox{ post}$&&$\widehat{\beta}\mbox{ prior}$&$\widehat{\beta}\mbox{ post}$\\\hline
0.1   && 0.007  &  0.008 && 0.101  &  0.101 && 0.007  &  0.008\\
0.2   && 0.006  &  0.007 && 0.199  &  0.199 && 0.006  &  0.007\\
0.3   && 0.006  &  0.006 && 0.302  &  0.301 && 0.005  &  0.006\\
0.4   && 0.006  &  0.007 && 0.404  &  0.403 && 0.005  &  0.006\\
0.45  && 0.006  &  0.006 && 0.454  &  0.454 && 0.004  &  0.004\\
0.5   && 0.004  &  0.004 && 0.502  &  0.502 && 0.003  &  0.004\\
0.6   &&  0.093  &  0.103 && 0.682   &  0.694 && 0.043  &  0.043\\
0.7   &&  0.031  &  0.034 && 0.719   &  0.723 && 0.024  &  0.025\\
0.8   &&  0.023  &  0.022 && 0.779   &  0.780 && 0.010  &  0.010\\
0.9   &&  0.043  &  0.039 && 0.858   &  0.863 && 0.013  &  0.015\\
1     &&  0.070  &  0.064 && 0.932   &  0.938 && 0.018  &  0.020\\  \bottomrule
\end{tabular}
\end{table}

\section{Sensitivity to model deviations}\label{Sec:Sensitivity}

In the previous section, we evaluated common accuracy measures when the estimation was produced over images generated by the true model.
This is never the case in practice.
Even if the radiometric data are emitted after a realization of the Potts model, such map remains unobserved.
As in all Hidden Markov Models, one of the main goals is to predict the most probable state label that could have emitted the observations.
Maximum likelihood classification generates an initial map, and ICM predicts the state map under the Potts prior.
Nevertheless an estimation of $\beta$ is needed, and it has to be computed from the initial classification, ML, or from a cycle of ICM made with an initial arbitrary $\beta$.

In this section, we explore the performance of the estimators in the setting where the map of classes is not an accurate estimation of a true Potts model realization.
Our study will involve the influence of the parameters of the emission in the curves $f_{\mbox{\scriptsize prior}}$ and  $f_{\mbox{\scriptsize post}}$, and in our estimations.

\subsection{Maximum Likelihood Classification as state map for estimation}

From the standard definition of numerical analysis, we will say that  a system of two linear equations with two unknowns is ill conditioned if the slopes of the equations are similar.
Ill conditioning makes that any small numerical error that displaces slightly the line changes greatly the point of intersection of the two lines.
Then, if a curve has a derivative with small absolute value, any displacement (even the smallest) moves the roots by a considerable amount.

In our context, the accuracy of the estimators as positions of the roots will be extremely dependent of the absolute value of the derivative of the $f_{\mbox{\scriptsize prior}}$ and  $f_{\mbox{\scriptsize post}}$. We compute such values in the following proposition.

\begin{proposition}
The derivatives of the functions defined in equations (\ref{f_prior}) and (\ref{f_post}) are
\begin{align}
\frac{d}{d\beta}f_{\mbox{\scriptsize prior}}(\beta)&=-\sum_{s\in S}\operatorname{Var}_{\beta}\left[U_s(X_s)\mid X_{\partial_s}\right]
\label{derivada_prior}\\
\frac{d}{d\beta}f_{\mbox{\scriptsize post}}(\beta)&=-\sum_{s\in S}\operatorname{Var}_{\beta}\left[U_s(X_s)\mid X_{\partial_s},I_s\right].
\label{derivada_post}
\end{align}
where the variances are computed respect to the conditional distribution $X_s\mid X_{\partial_s}$ and $X_s\mid (X_{\partial_s},I_s)$ respectively.
\end{proposition}

The derivatives from both curves coincide when both classes emit observations $I_s$ under the same distribution.
As the means grow apart, the absolute value of the slope of $f_{\mbox{\scriptsize post}}$ decreases, intersecting the $\beta$ axis at smaller angles, which in turn makes the estimation more sensitive to numerical errors.
In the case of the $f_{\mbox{\scriptsize prior}}$ curve, the distance between the means has no influence on the root positions.
To differentiate this curves from the curves generated by using ML data, we define the following functions
\begin{align}
f_{\mbox{\scriptsize prior}}^{\mbox{\scriptsize ML}}(\beta)=&\displaystyle\sum_{s\in S}U_s^{\mbox{\scriptsize ML}}(\widehat{\bm x}_{\mbox{\scriptsize ML},s})-\nonumber\\
&-\sum_{s\in S}\frac{\sum_{\ell\in\mathcal{L}}U_s^{\mbox{\scriptsize ML}}(\ell)\exp\{\beta U_s^{\mbox{\scriptsize ML}}(\ell)\}}{\sum_{\ell\in\mathcal{L}}\exp\{\beta U_s^{\mbox{\scriptsize ML}}(\ell)\}}
\label{f_priorML}
\end{align}
and
\begin{align}
f_{\mbox{\scriptsize post}}^{\mbox{\scriptsize ML}}(\beta)=&\displaystyle\sum_{s\in S}U_s^{\mbox{\scriptsize ML}}(\widehat{\bm x}_{\mbox{\scriptsize ML},s})-\nonumber\\
&-\sum_{s\in S}\frac{\sum_{\ell\in\mathcal{L}}U_s^{\mbox{\scriptsize ML}}(\ell)p(I_s\mid \ell)\exp\{\beta U_s^{\mbox{\scriptsize ML}}(\ell)\}}{\sum_{\ell\in\mathcal{L}}p(I_s\mid \ell)\exp\{\beta U_s^{\mbox{\scriptsize ML}}(\ell)\}},
\label{f_postML}
\end{align}
where $U_s^{\mbox{\scriptsize ML}}(\ell)=\#\{t\in\partial_s:\widehat{\bm x}_{\mbox{\scriptsize ML},t}=\ell\}$ and $\widehat{\bm x}_{\mbox{\scriptsize ML},s}$ is the state value in the pixel $s$  in the ML classification of $\bm I$.

\begin{figure}[h!]
\centering
\subfigure[$k=1$]{
  \includegraphics[scale = 0.24]{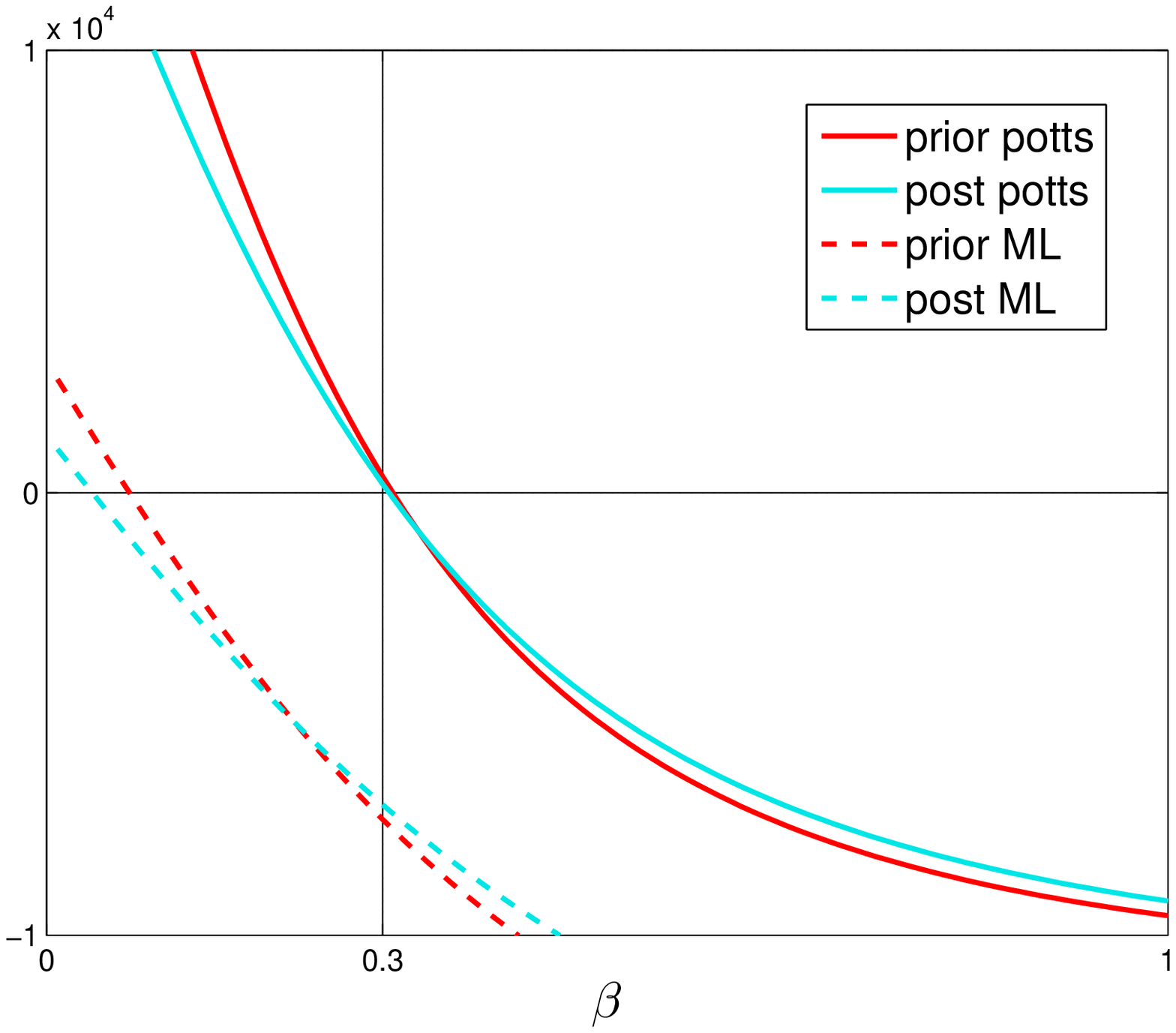}
}
\subfigure[$k=2$]{
  \includegraphics[scale = 0.24]{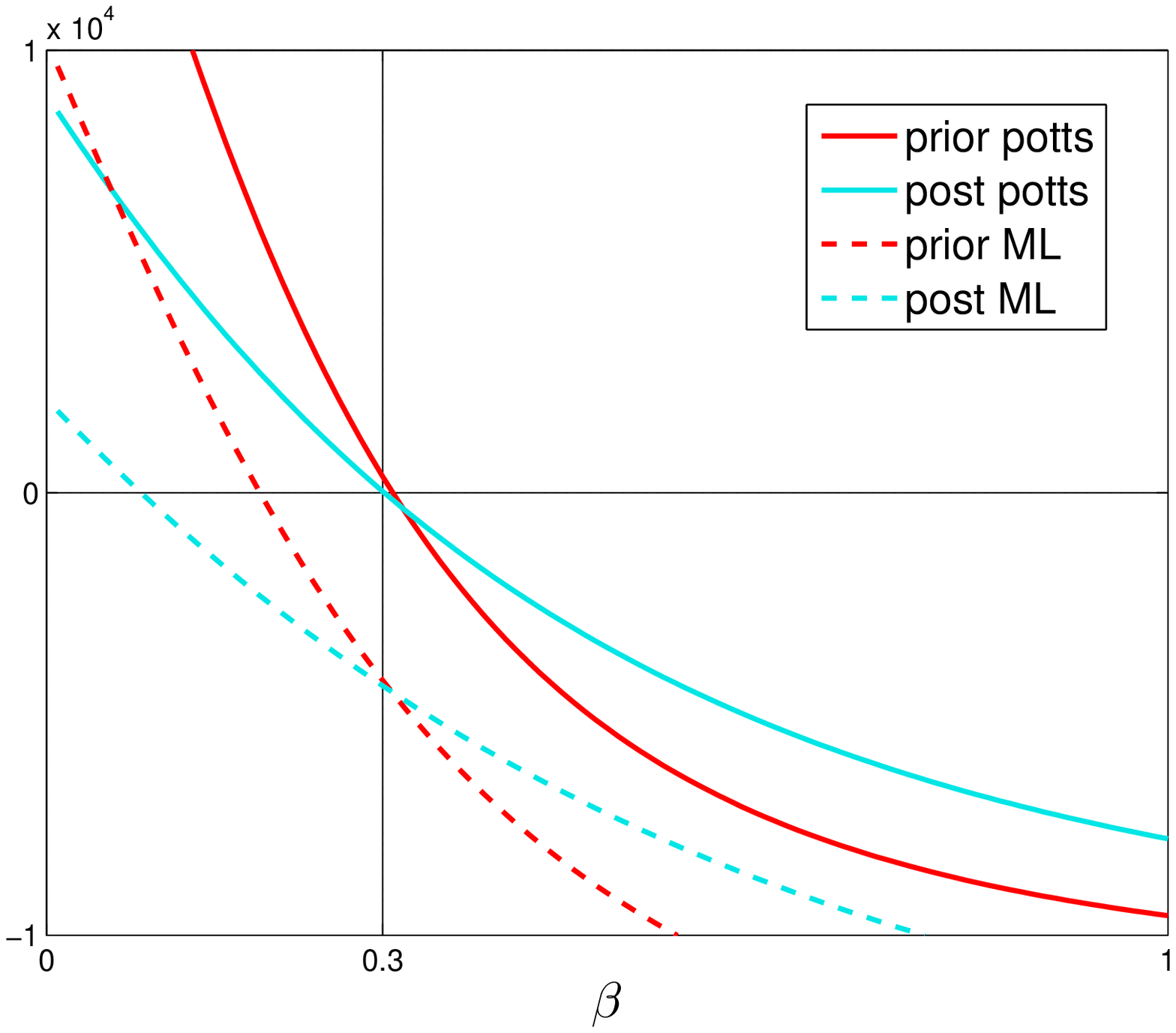}
}
\subfigure[$k=3$]{
  \includegraphics[scale = 0.24]{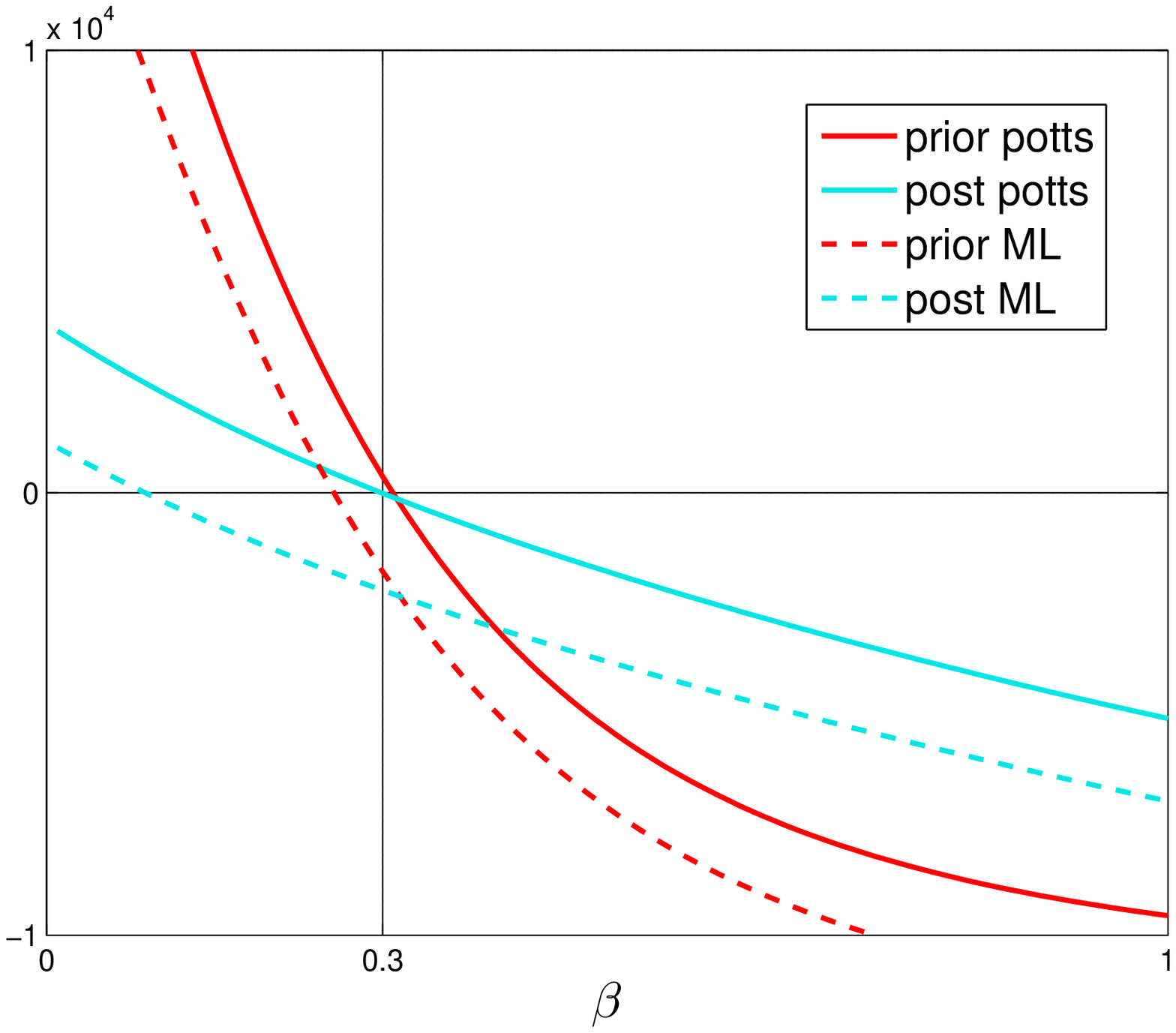}
}
\subfigure[$k=4$]{
  \includegraphics[scale = 0.24]{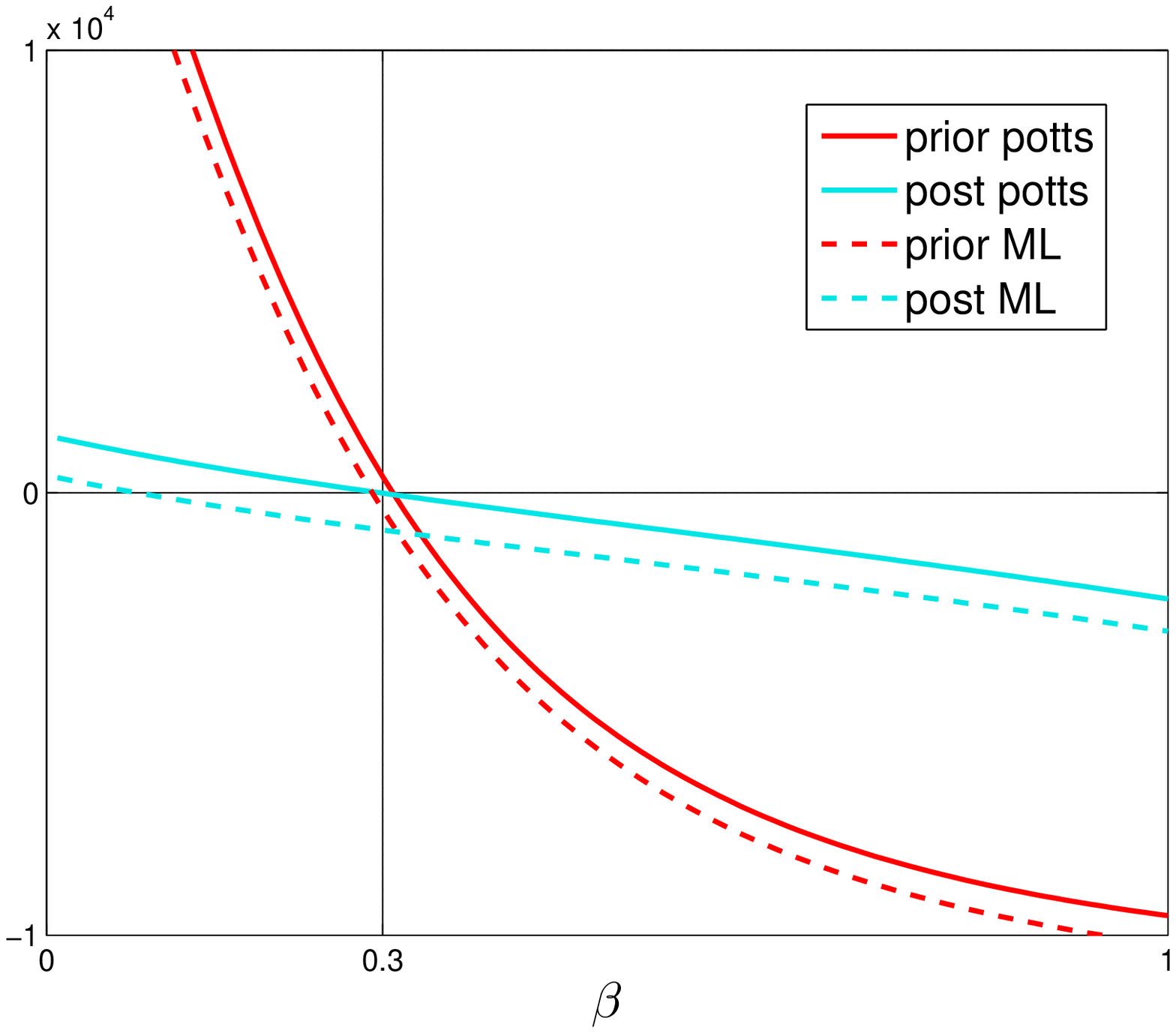}
}
\caption{Plots of $f_{\mbox{\scriptsize prior}}$, $f_{\mbox{\scriptsize post}}$, $f_{\mbox{\scriptsize prior}}^{\mbox{\scriptsize ML}}$ and $f_{\mbox{\scriptsize post}}^{\mbox{\scriptsize ML}}$ for $L=2$; $\beta=0.3$; and several values of $k$.}
 \label{grafico_muestra_sensibilidad}
\end{figure}

In each of the plots of Fig.~\ref{grafico_muestra_sensibilidad} the sensibility of the estimators is illustrated for different lags (differences between means), $L=2$  and $\beta=0.3$.
The plots present in red and cyan the curves corresponding to the prior and posterior model, respectively.
This curves appear in full lines if they pertain to the pure model ($f_{\mbox{\scriptsize prior}}$ and $f_{\mbox{\scriptsize post}}$), and in dashed lines if they pertain to the contaminated model with the ML segmentations ($f_{\mbox{\scriptsize prior}}^{\mbox{\scriptsize ML}}$ and $f_{\mbox{\scriptsize post}}$).

The curves of the prior statistic under the pure model are very similar in all the plots, since $k$ does not affect  $f_{\mbox{\scriptsize prior}}$.
In the other hand, curves of the posterior statistic under the pure model show the incidence of $k$, this is, as the lag increases, the curve tends to an horizontal line.
Nevertheless, under the pure model, both statistics present good and indistinguishable estimations.

The curves of the prior statistic under the contaminated model are almost parallel to the curves corresponding to the pure model, but their roots are smaller than the true $\beta$.
The larger the lags the better the segmentation produced by ML, i.e., closest to the true Potts model realization, and the better the estimation producing by the $f_{\mbox{\scriptsize prior}}^{\mbox{\scriptsize ML}}$ curve root.
Despite the displacement in the curves being the same for both estimators, the estimator based on the posterior model is more influenced, having a curve with smaller slope.

From this, two complementary concepts arise when the lags increase: parallel curves of the same color are closer, and estimations with larger negative bias produced by the reduction in the slope of the curves.
This two concepts are mixed in the posterior estimator leading to no improvement when the lags increase.
For the prior estimator, separation between the class conditional distributions improves the estimation.

We should note that $f_{\mbox{\scriptsize prior}}^{\mbox{\scriptsize ML}}\leq f_{\mbox{\scriptsize prior}}$ and that $f_{\mbox{\scriptsize post}}^{\mbox{\scriptsize ML}}\leq f_{\mbox{\scriptsize post}}$.
This follows form the fact the the expressions are regulated by their first term.
Such term is larger in the pure model since the maps are more homogeneous than the ones estimated by ML.
This implies the estimations being smaller, thus biased to the left.

\begin{figure}[h!]
\centering
\subfigure[$k=1$\label{fig_haces_ML_1}]{
  \includegraphics[scale = 0.24]{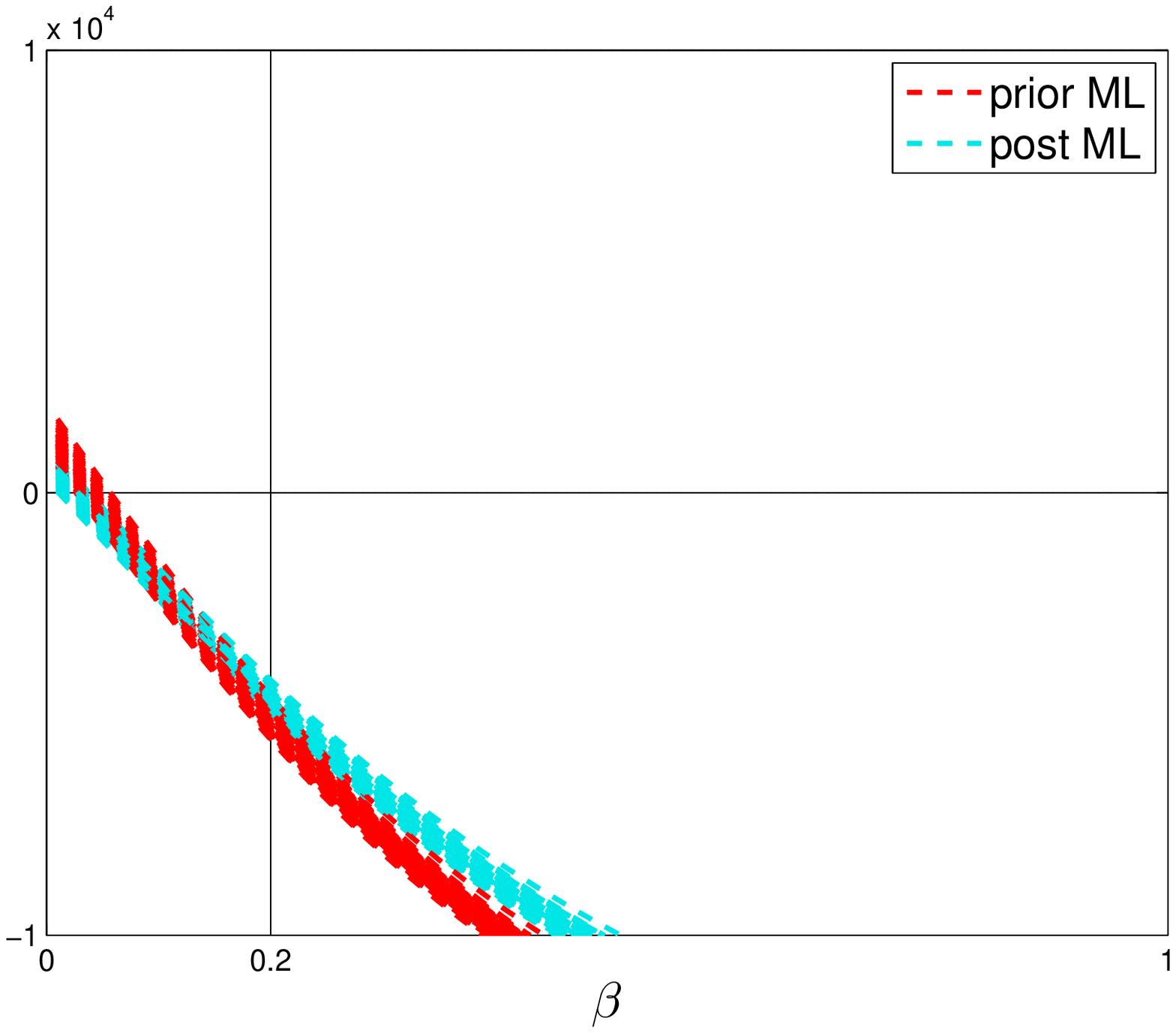}
}
\subfigure[$k=4$\label{fig_haces_ML_4}]{
  \includegraphics[scale = 0.24]{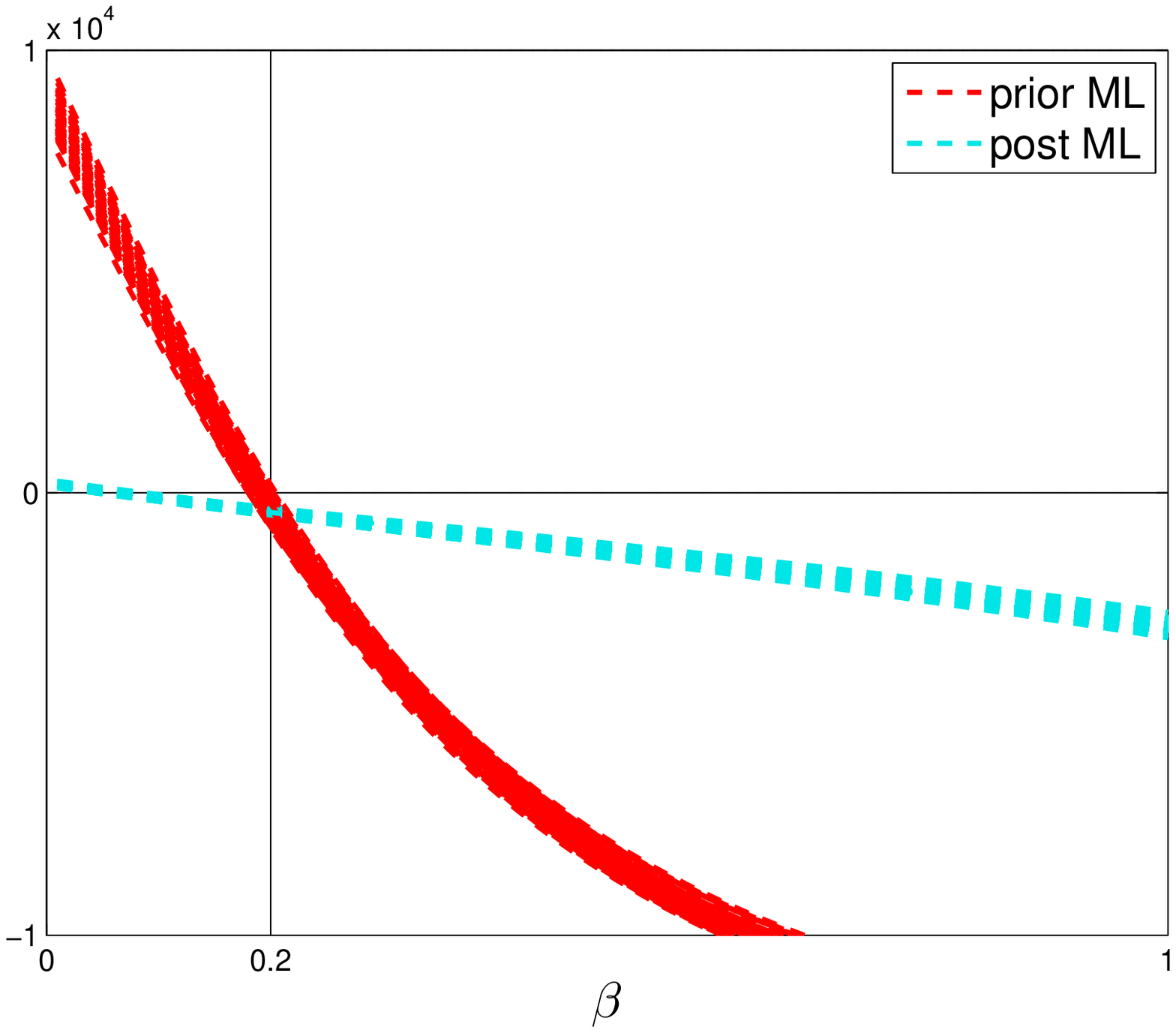}
}
\caption{Bundles of curves of the functions $f_{\mbox{\scriptsize prior}}^{\mbox{\scriptsize ML}}$ and $f_{\mbox{\scriptsize post}}^{\mbox{\scriptsize ML}}$ for $L=2$; $\beta=0.2$; and $k\in\{1,4\}$.}
\label{grafico_haz_ML}
\end{figure}

Fig.~\ref{grafico_haz_ML} shows, as well than Fig.~\ref{fig_haces_fprior_fpost}, the bundles of curves of both estimators under the contaminated model, for the case $L=2$,  $\beta=0.2$, and $k=1,4$.

\begin{figure*}[htb]
 \[\begin{array}{@{}ccc@{}} \toprule
L=2 &L=3&L=4\\ \hline
&k=1&\\ \hline
\includegraphics[scale= 0.25]{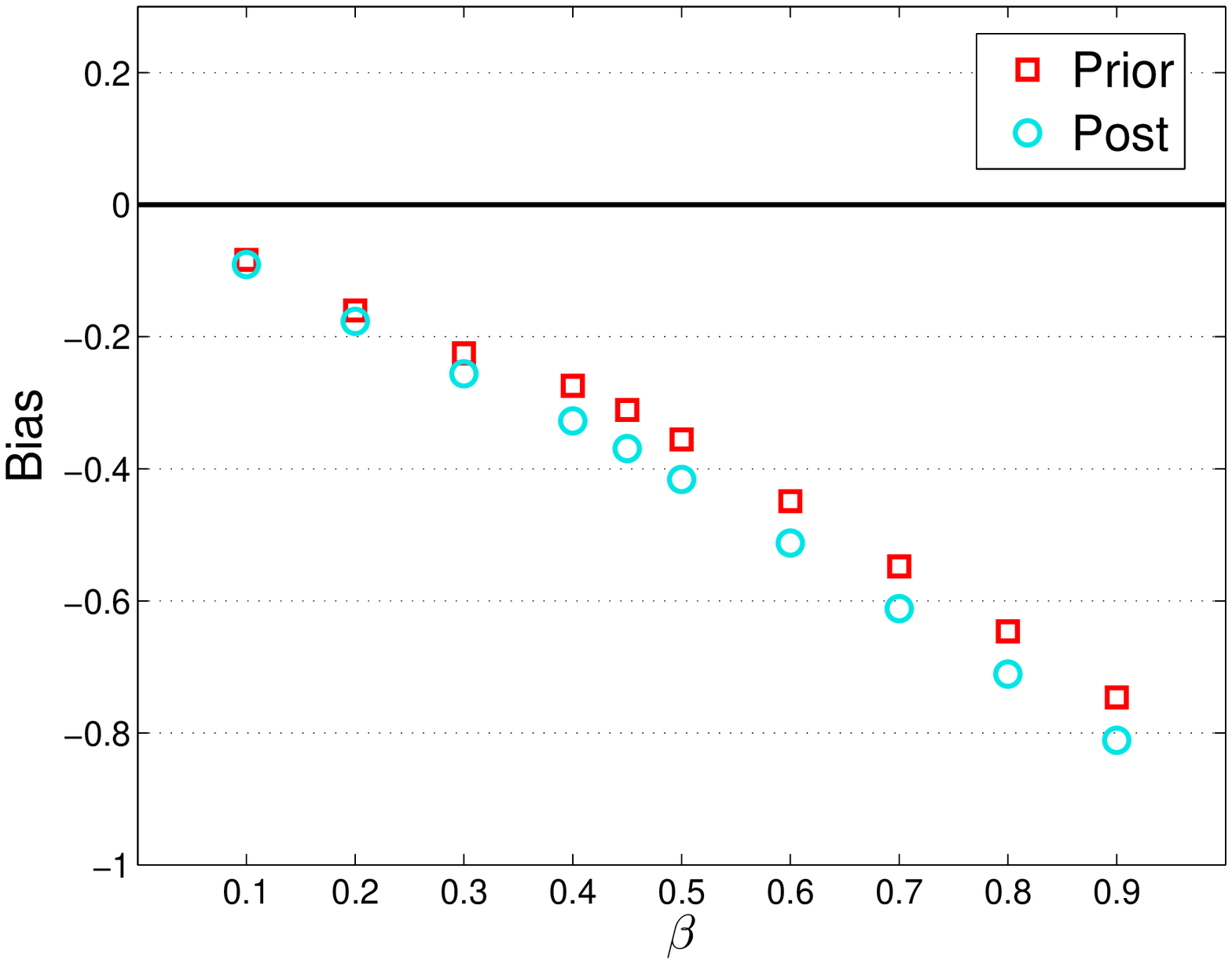}&\includegraphics[scale= 0.25]{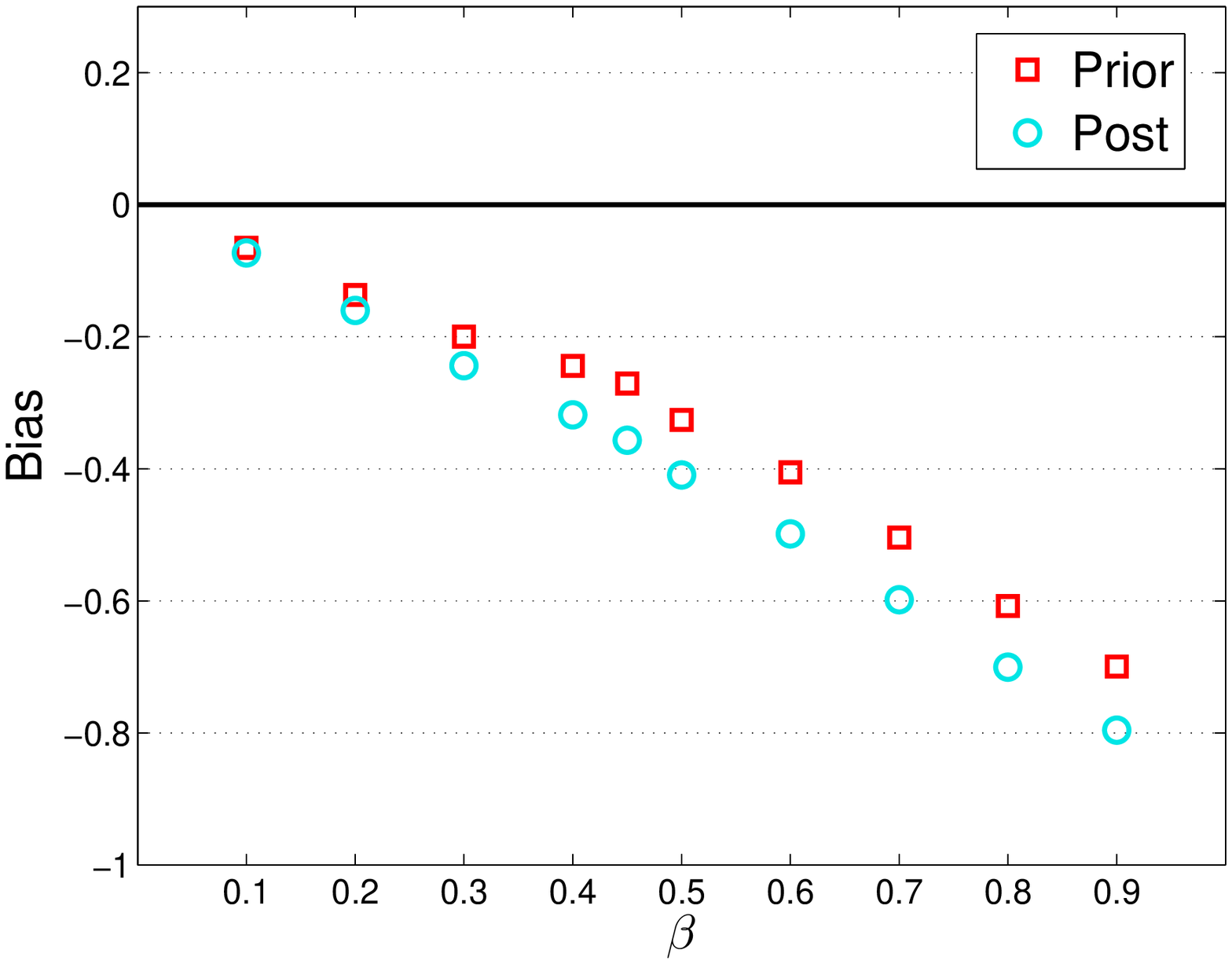}&
\includegraphics[scale= 0.25]{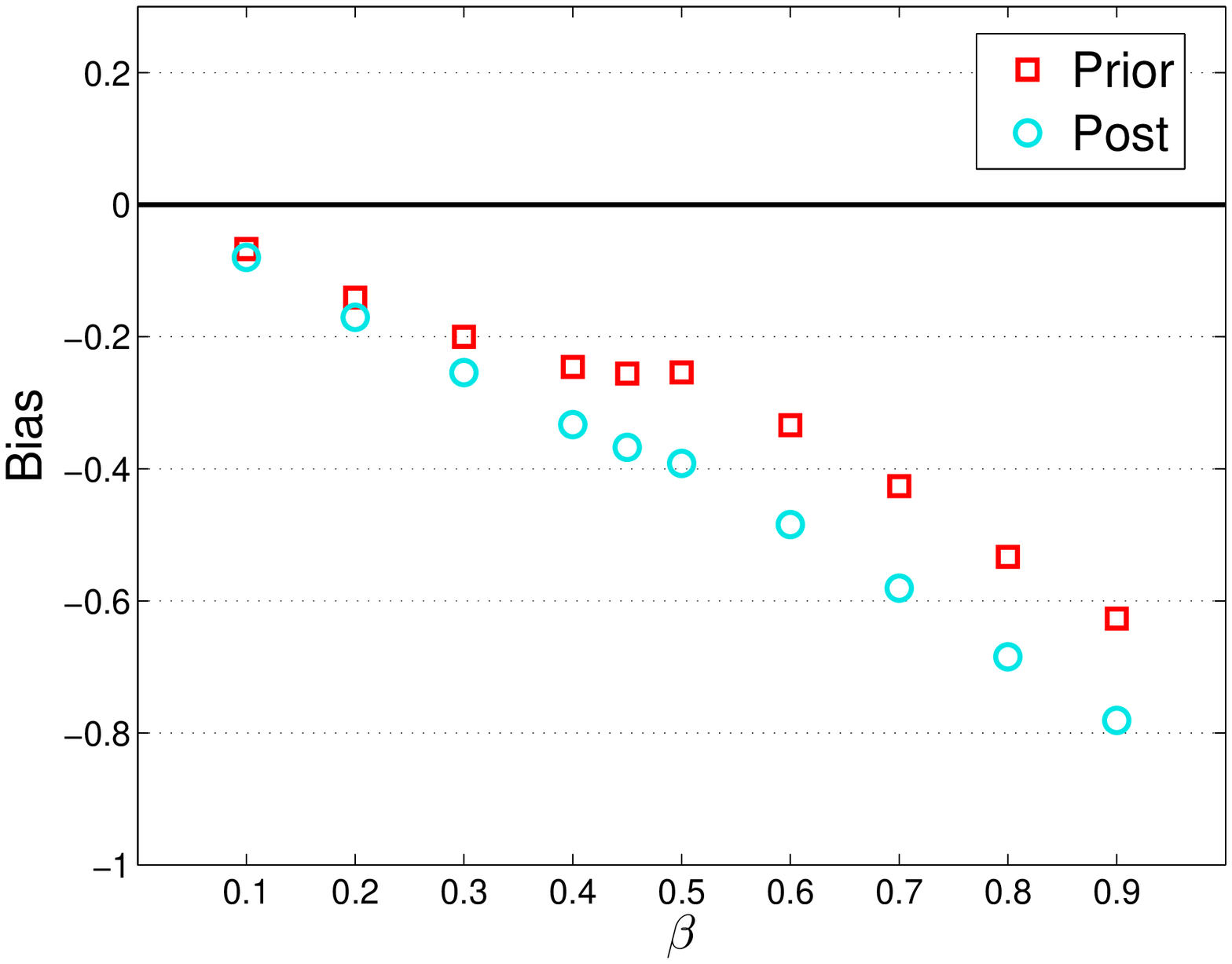}\\ \hline
&k=4&\\ \hline
\includegraphics[scale= 0.25]{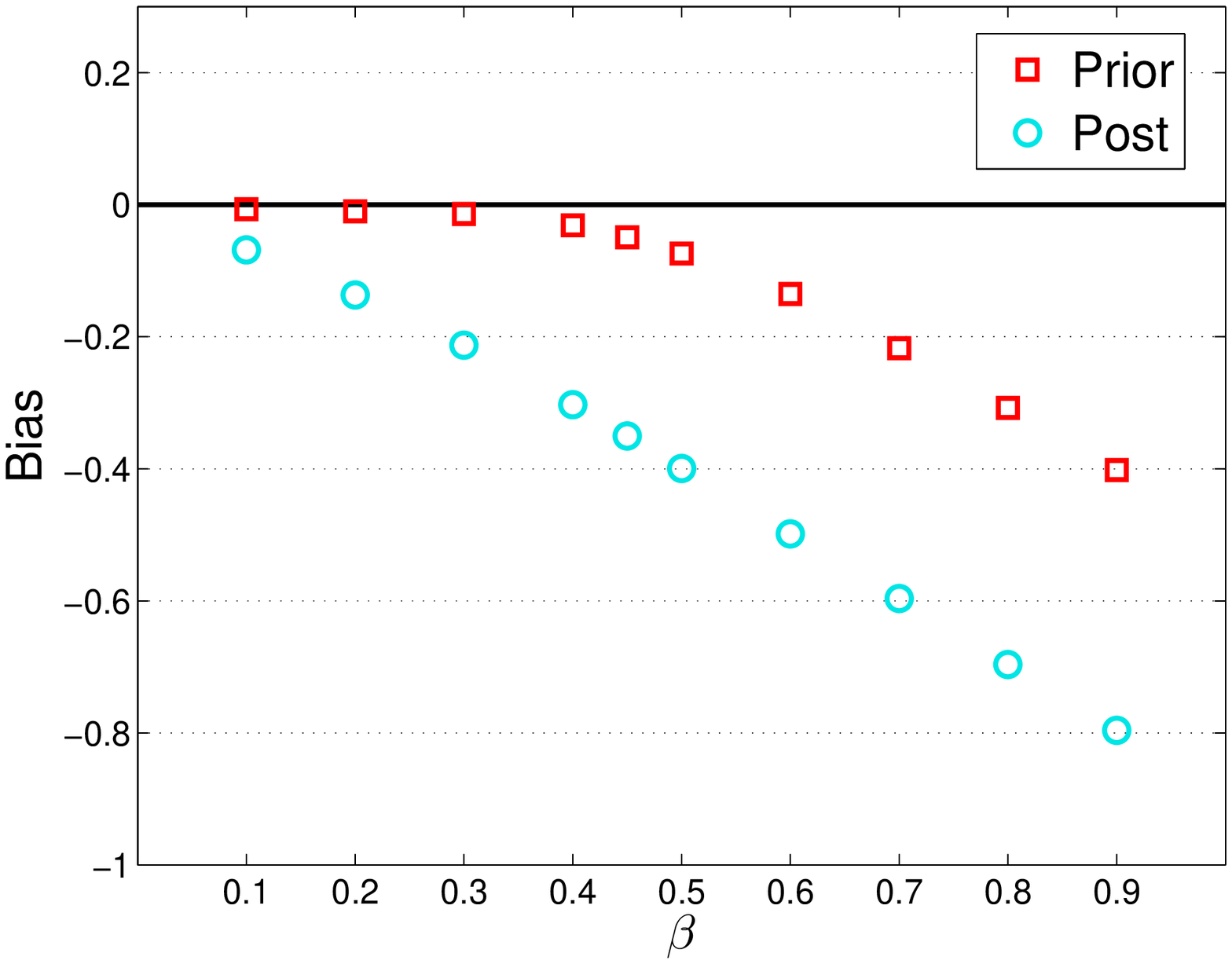}&\includegraphics[scale= 0.25]{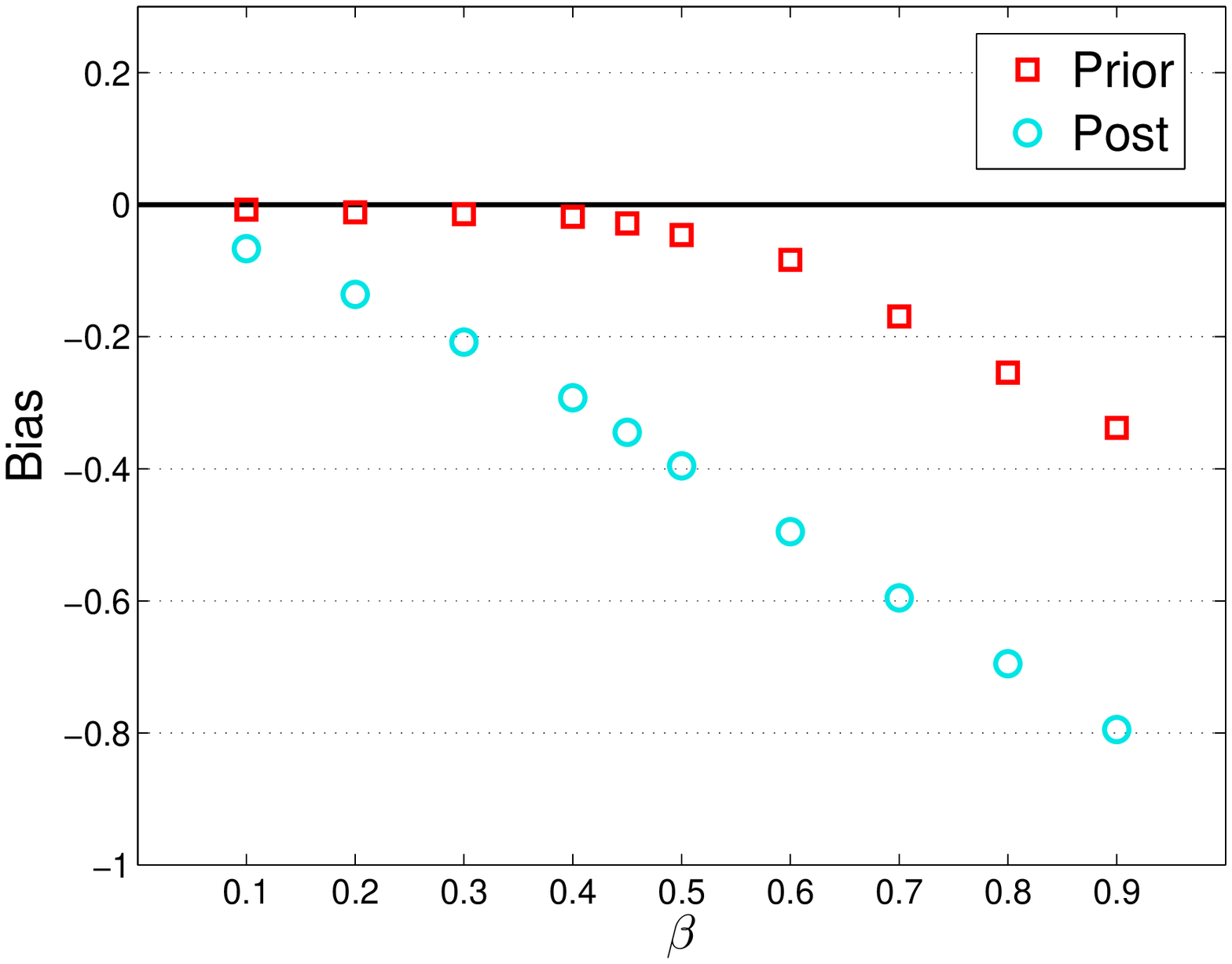}&
\includegraphics[scale= 0.25]{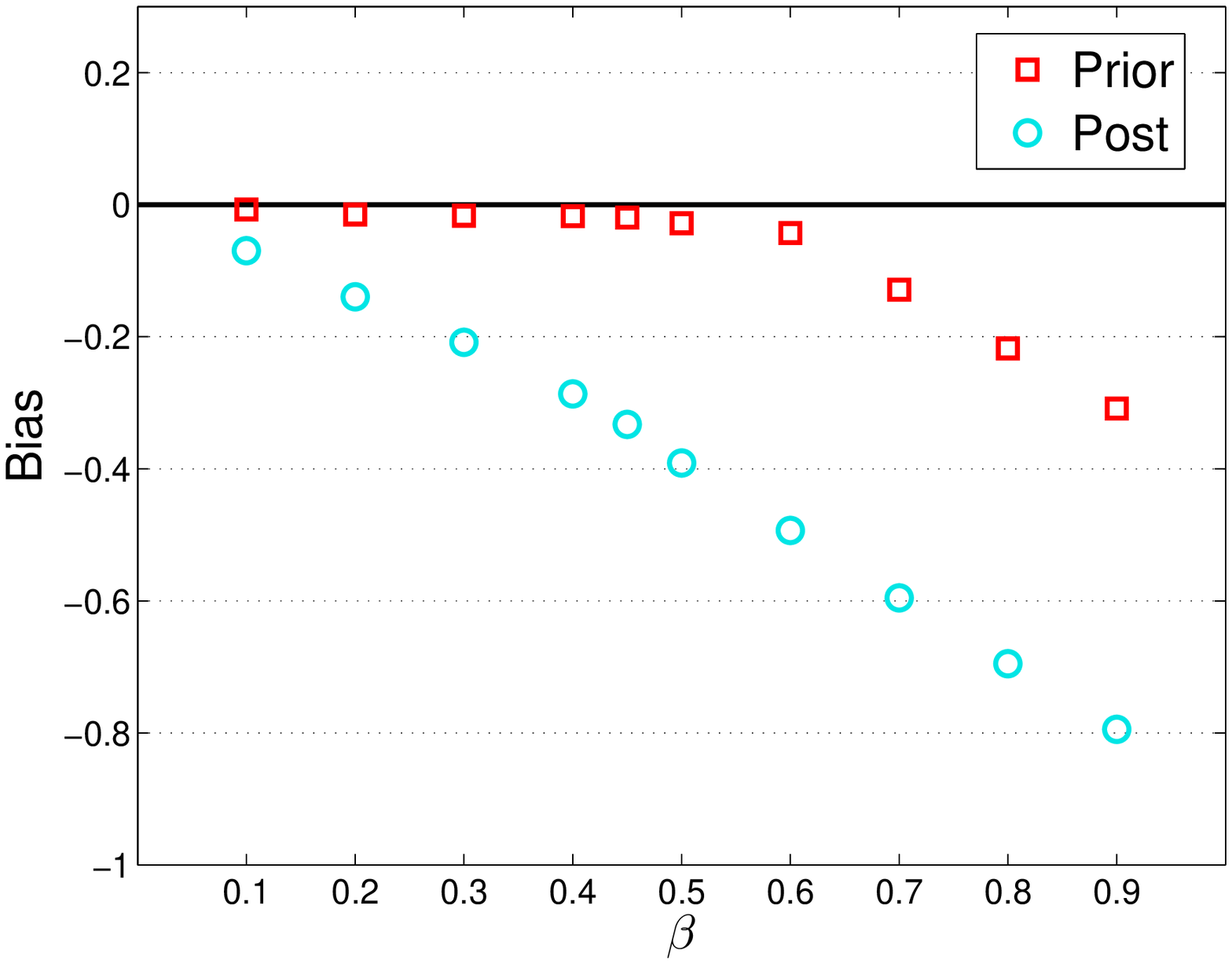}\\ \bottomrule
 \end{array} \]
\caption{Bias of the estimators produced by using $\widehat{\bm x}_{\mbox{\scriptsize ML}}$ instead of the Potts model realization, for $L\in\{2,3,4\}$; $k\in\{1,4\}$; and several values of $\beta$.}
 \label{grafico_ML_bias_vs_beta}
\end{figure*}

Fig.~\ref{grafico_ML_bias_vs_beta} shows the bias of both estimators as a function of $\beta$, for several $L$ and $k$, when the model is contaminated with the $\widehat{\bm x}_{\mbox{\scriptsize ML}}$ estimated map.
Clearly, both estimators show a marked bias to the left, reduced only in the case of the prior estimator when $k$ increases and $\beta$ has values smaller than $0.5$.

\subsection{Iterated conditional Mode as class map for estimation}
\begin{figure}[htb]
\centering
\subfigure[$k=1$]{
  \includegraphics[scale = 0.24]{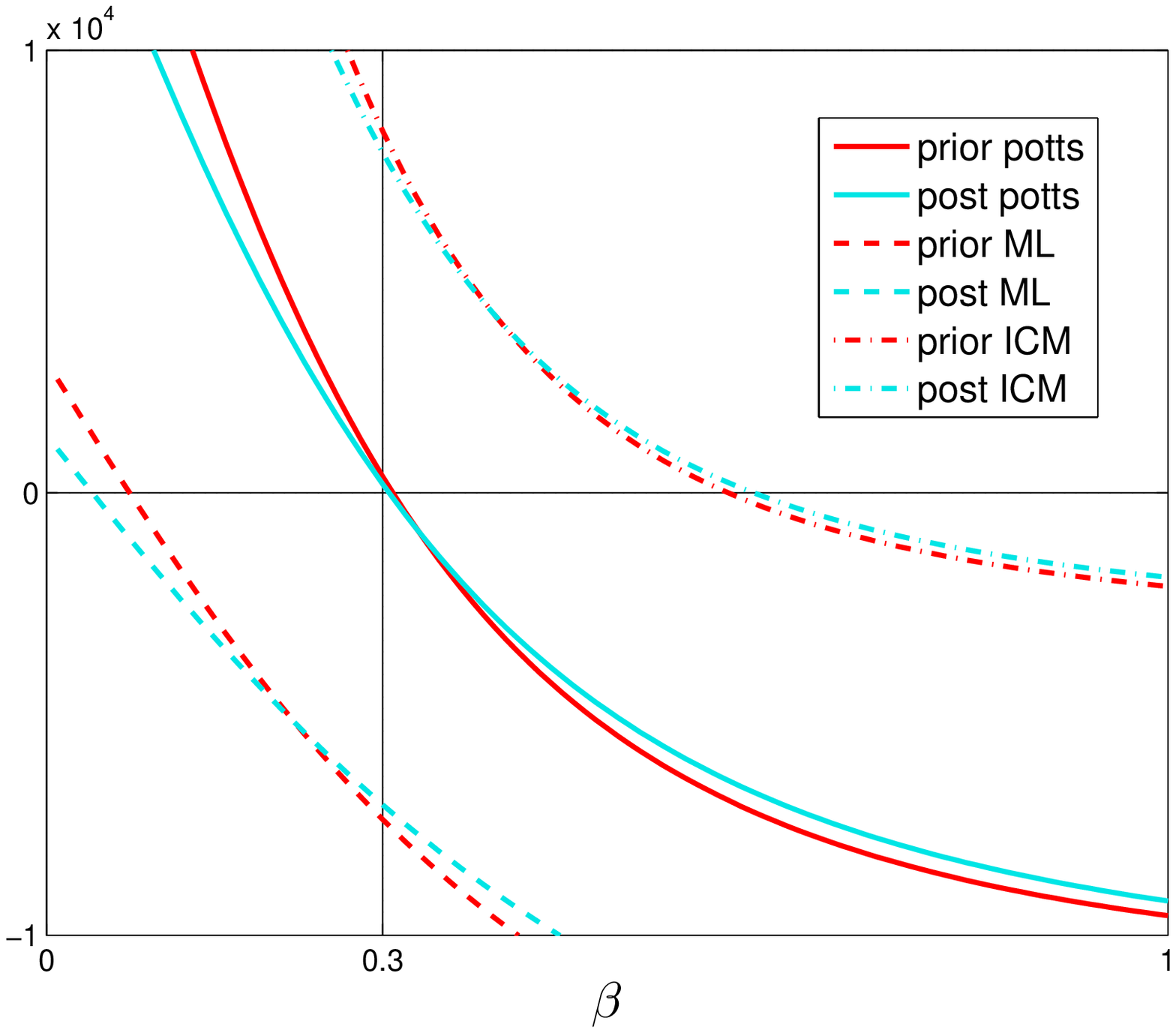}
  \label{fig_sensICM_1}
}
\subfigure[$k=2$]{
  \includegraphics[scale = 0.24]{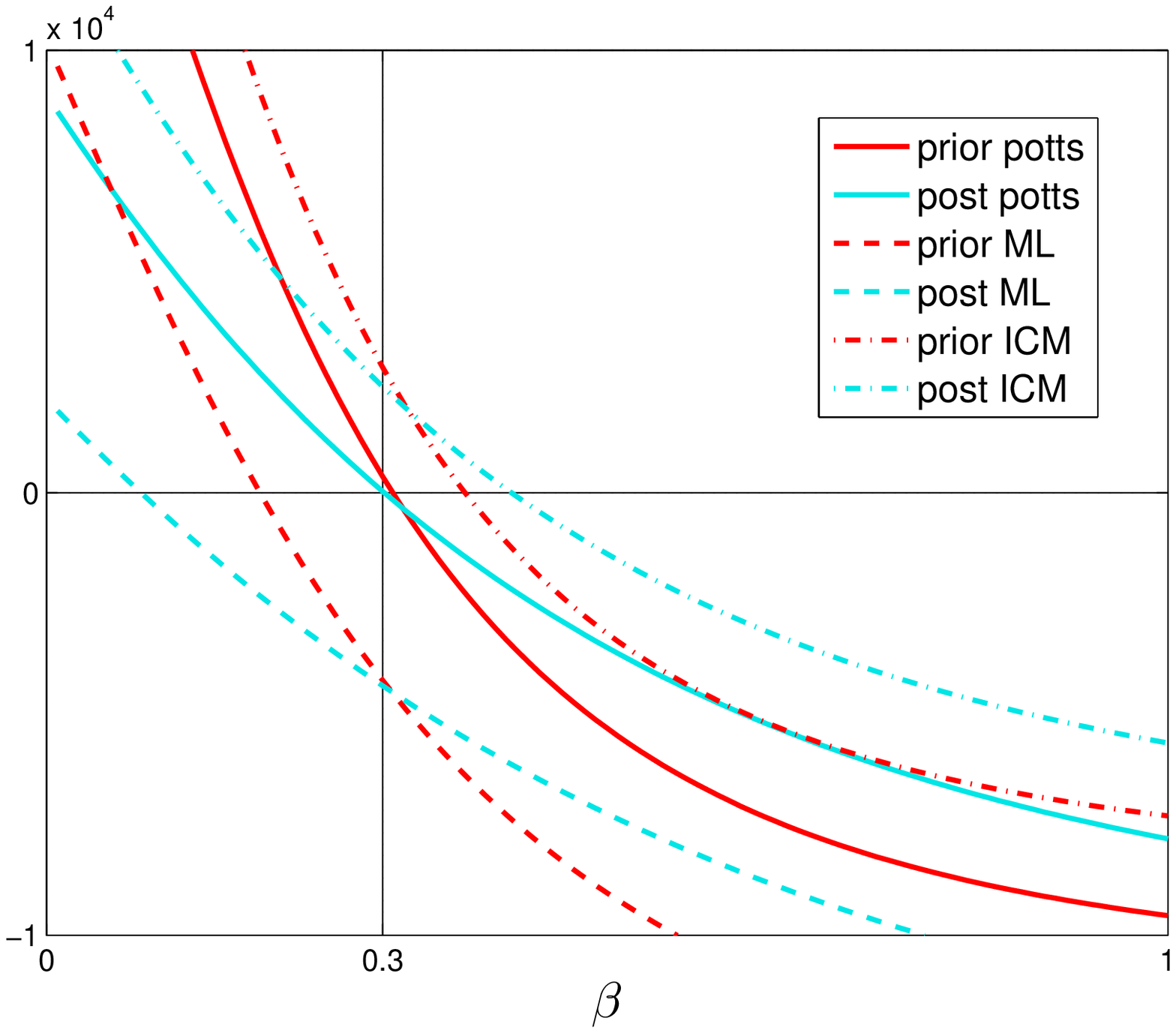}
  \label{fig_sensICM_2}
}
\subfigure[$k=3$]{
  \includegraphics[scale = 0.24]{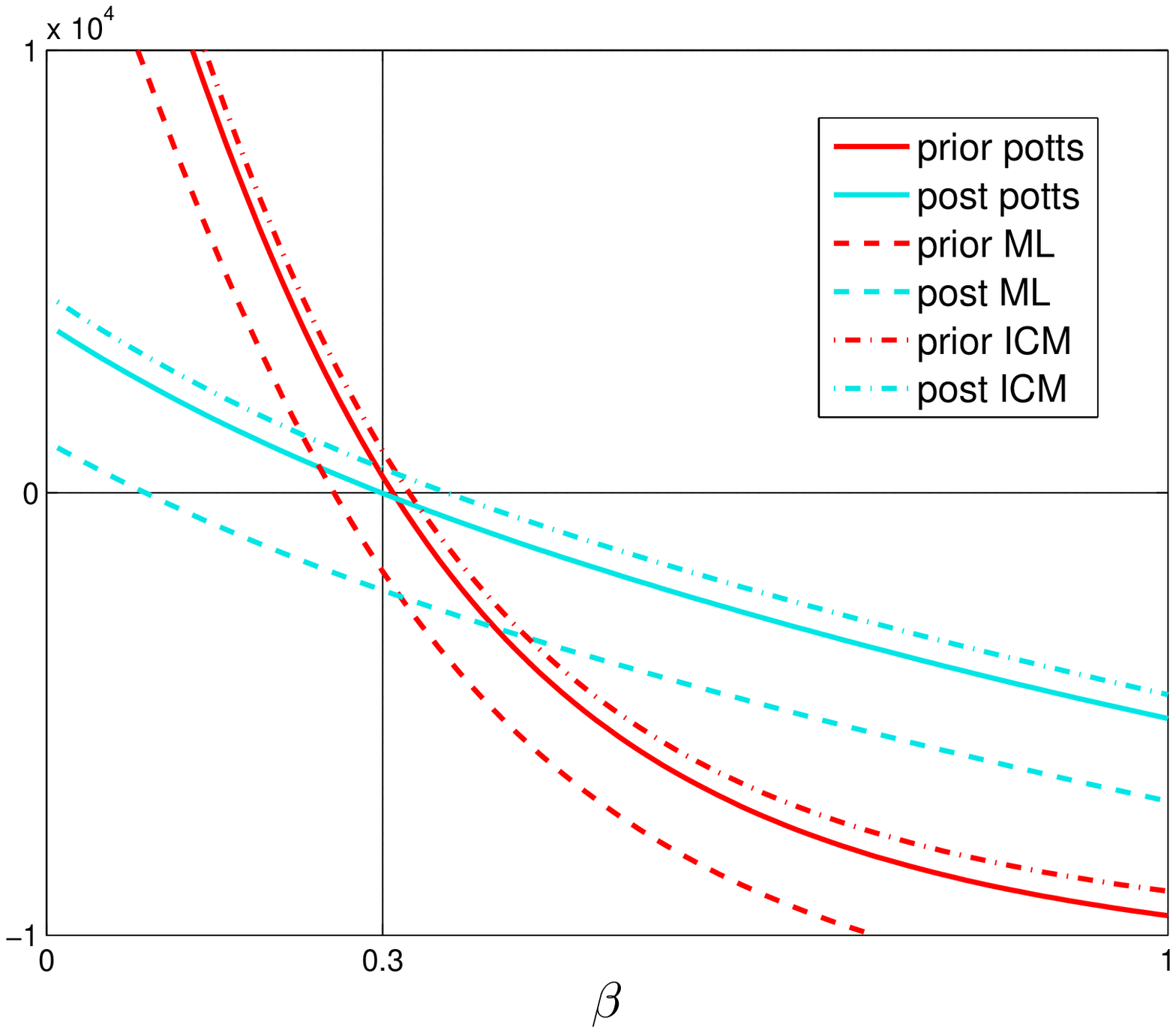}
  \label{fig_sensICM_3}
}
\subfigure[$k=4$]{
  \includegraphics[scale = 0.24]{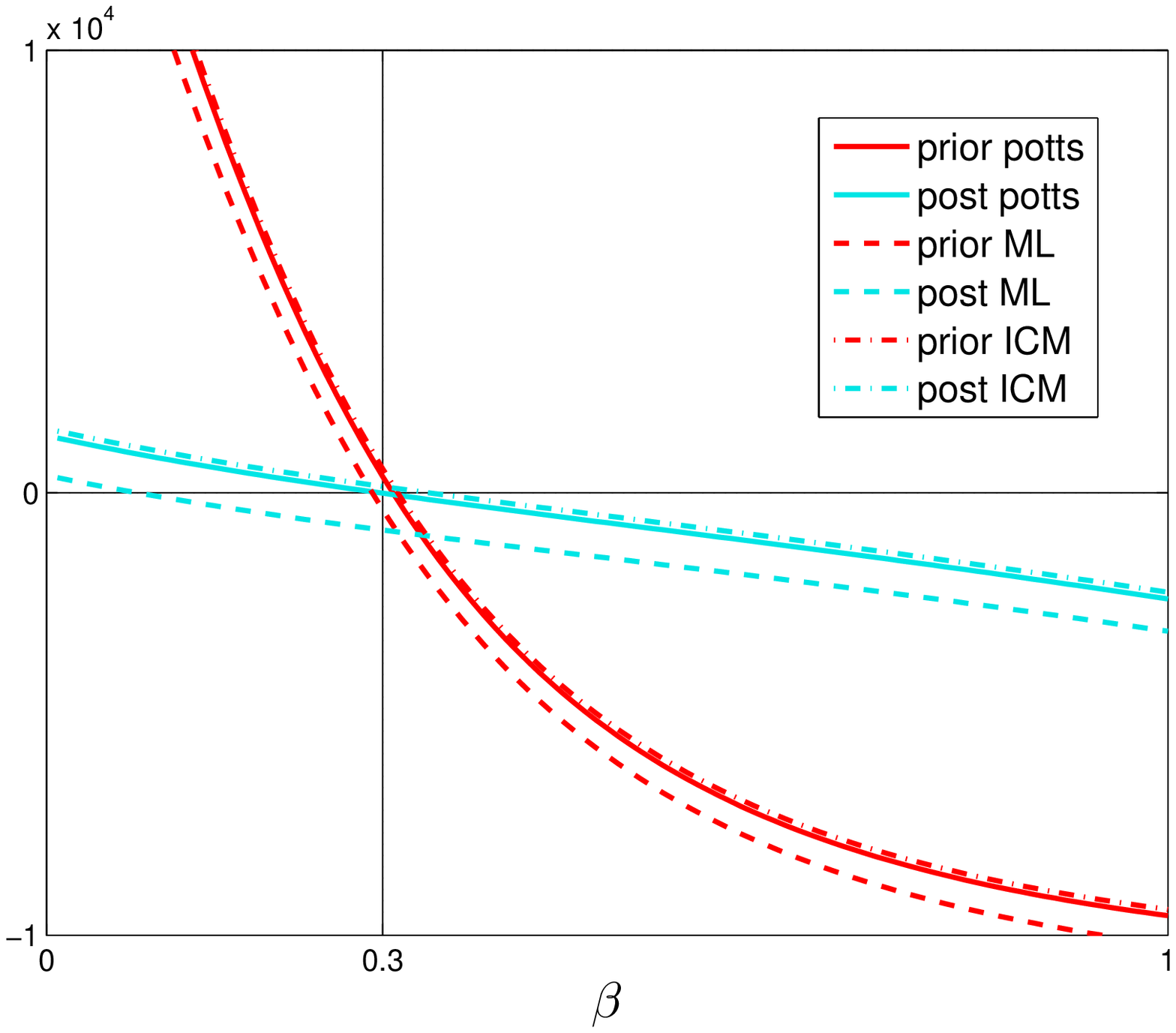}
  \label{fig_sensICM_4}
}
\caption{Plots of $f_{\mbox{\scriptsize prior}}$, $f_{\mbox{\scriptsize post}}$, $f_{\mbox{\scriptsize prior}}^{\mbox{\scriptsize ML}}$, $f_{\mbox{\scriptsize post}}^{\mbox{\scriptsize ML}}$, $f_{\mbox{\scriptsize prior}}^{\mbox{\scriptsize ICM}}$ and $f_{\mbox{\scriptsize post}}^{\mbox{\scriptsize ICM}}$ for $L=2$; $\beta=0.3$; and several values of $k$.}
 \label{grafico_muestra_sensibilidad_ICM}
\end{figure}

In a similar way, we define the functions $f_{\mbox{\scriptsize prior}}^{\mbox{\scriptsize ICM}}$ and $f_{\mbox{\scriptsize post}}^{\mbox{\scriptsize ICM}}$, using instead of $\widehat{\bm x}_{\mbox{\scriptsize ML}}$ the ICM output segmentation $\widehat{\bm x}_{\mbox{\scriptsize ICM}}$, setting $\beta$ as the parameter used in the simulation of the Potts model.
Fig.~\ref{grafico_muestra_sensibilidad_ICM} shows the plots of the functions $f_{\mbox{\scriptsize prior}}^{\mbox{\scriptsize ML}}$, $f_{\mbox{\scriptsize prior}}$ and $f_{\mbox{\scriptsize prior}}^{\mbox{\scriptsize ICM}}$ in red, and the functions $f_{\mbox{\scriptsize post}}^{\mbox{\scriptsize ML}}$, $f_{\mbox{\scriptsize post}}$ and $f_{\mbox{\scriptsize post}}^{\mbox{\scriptsize ICM}}$ in cyan, for different $k$.
The curves corresponding to the pure model are full lines, and the dashed lines and dotted lines are the curves corresponding to the contaminated model, when using ML and ICM as observed class map, $\widehat{\bm x}_{\mbox{\scriptsize ML}}$ and $\widehat{\bm x}_{\mbox{\scriptsize ICM}}$, respectively.


We see that  $f_{\mbox{\scriptsize prior}}^{\mbox{\scriptsize ML}}\leq f_{\mbox{\scriptsize prior}}^{\mbox{\scriptsize ICM}}$ and $f_{\mbox{\scriptsize post}}^{\mbox{\scriptsize ML}}\leq f_{\mbox{\scriptsize post}}^{\mbox{\scriptsize ICM}}$.
This due to the smoothing that ICM introduces over its initial map $\widehat{\bm x}_{\mbox{\scriptsize ML}}$, producing larger estimates than the ones computed directly over  $\widehat{\bm x}_{\mbox{\scriptsize ML}}$.

Besides, it holds that $f_{\mbox{\scriptsize prior}}\leq f_{\mbox{\scriptsize prior}}^{\mbox{\scriptsize ICM}}$ and $f_{\mbox{\scriptsize post}}\leq f_{\mbox{\scriptsize post}}^{\mbox{\scriptsize ICM}}$, due to the fact that $\widehat{\bm x}_{\mbox{\scriptsize ICM}}$ is a local maximum of equation (\ref{posteriori}), and, because of that, it is smoother than a random realization of the model  (\ref{posteriori}), which is not necessary a mode of such model.

Like before, we should note that the curves $f_{\mbox{\scriptsize post}}^{\mbox{\scriptsize ICM}}$ reduce curvature when $k$ increases.
In the other hand, while $k$ increases the curves of the contaminated models get closer to the curves of the pure model.
This is because the larger $k$ is, the more it will cost to ICM generate connected regions in the map since in such case the contextual evidence is weaker than the radiometric evidence.

\begin{figure*}[htb]
 \[\begin{array}{ccc} \toprule
L=2 &L=3&L=4\\ \hline
&k=1&\\ \hline
\includegraphics[scale= 0.25]{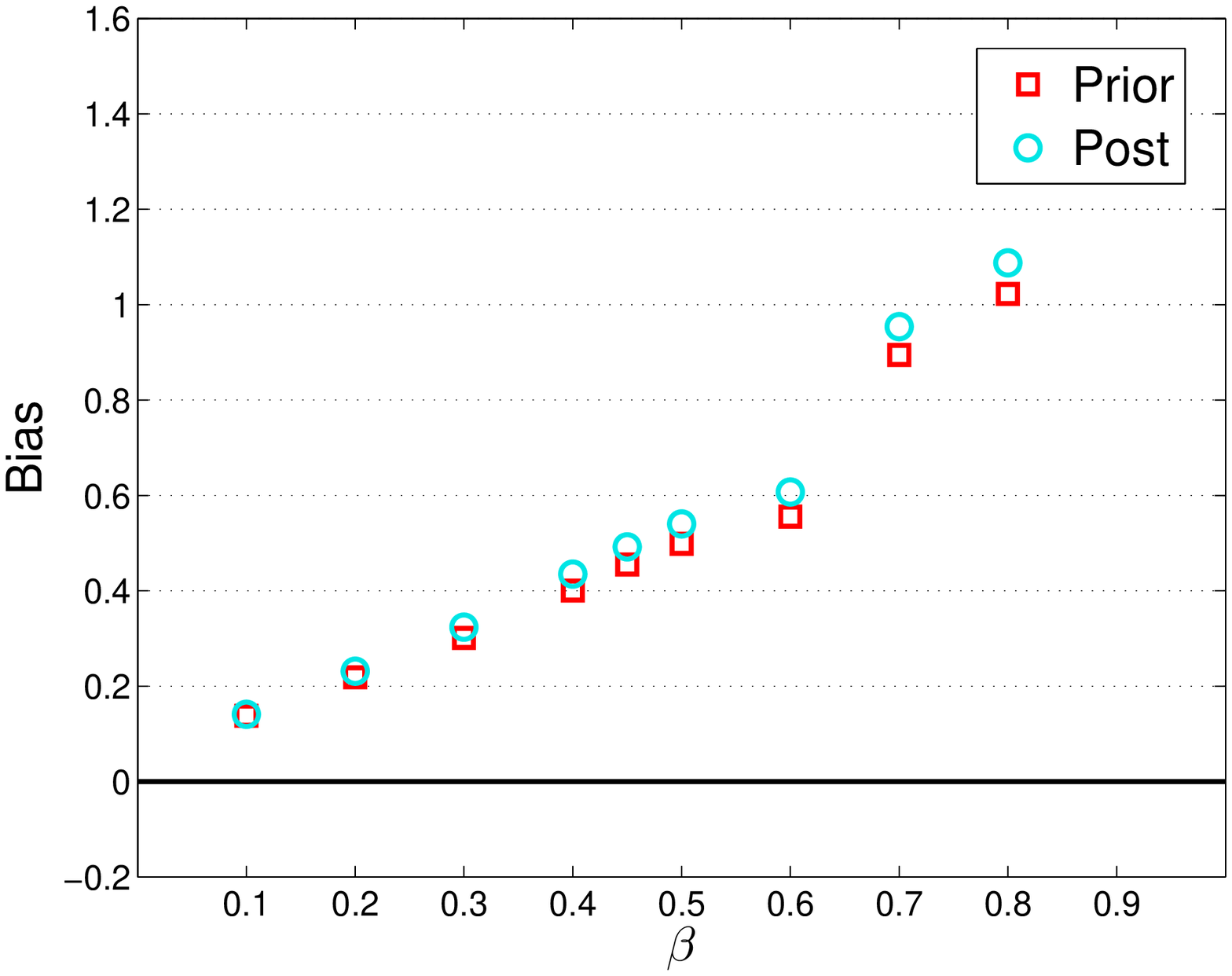}&\includegraphics[scale= 0.25]{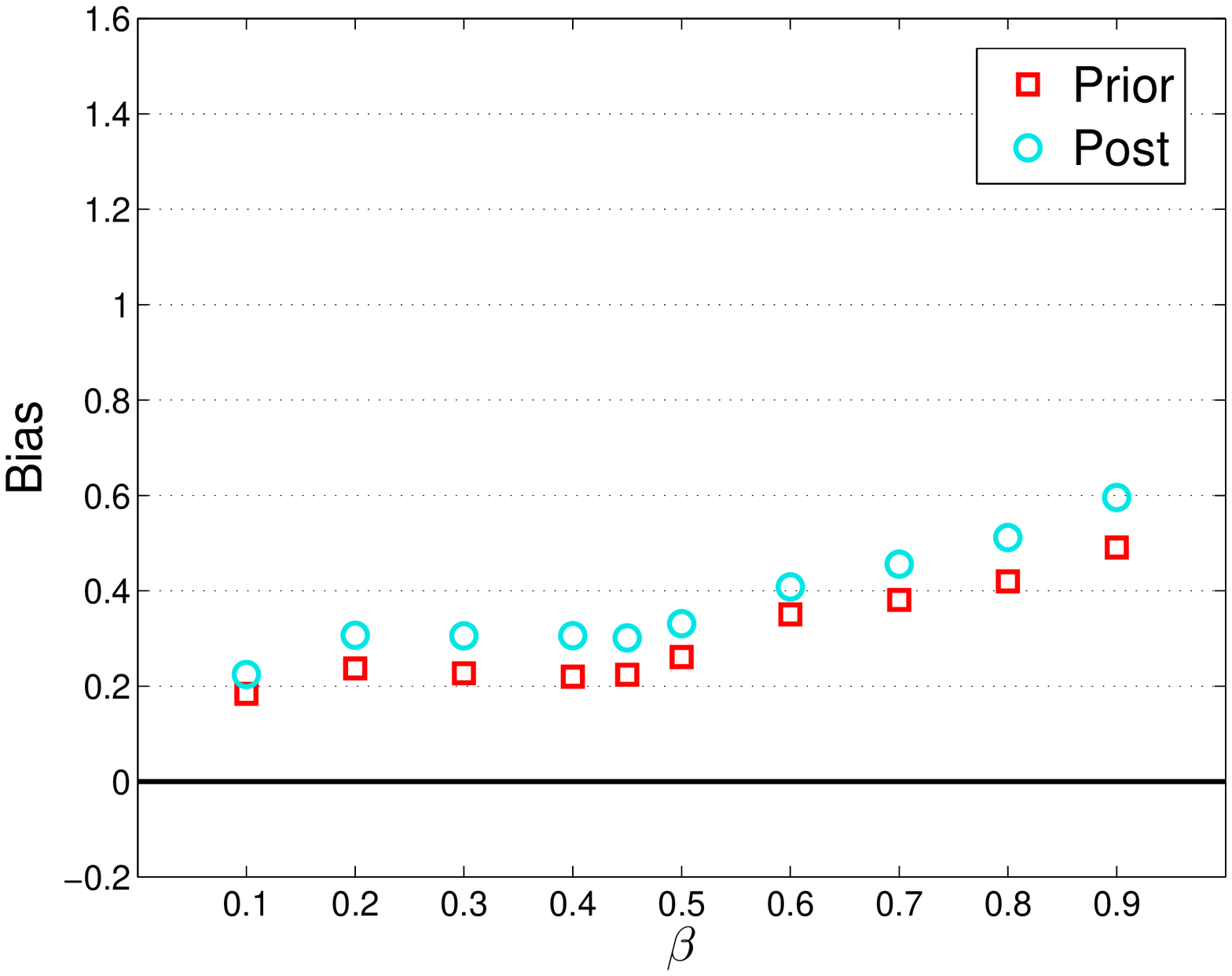}&
\includegraphics[scale= 0.25]{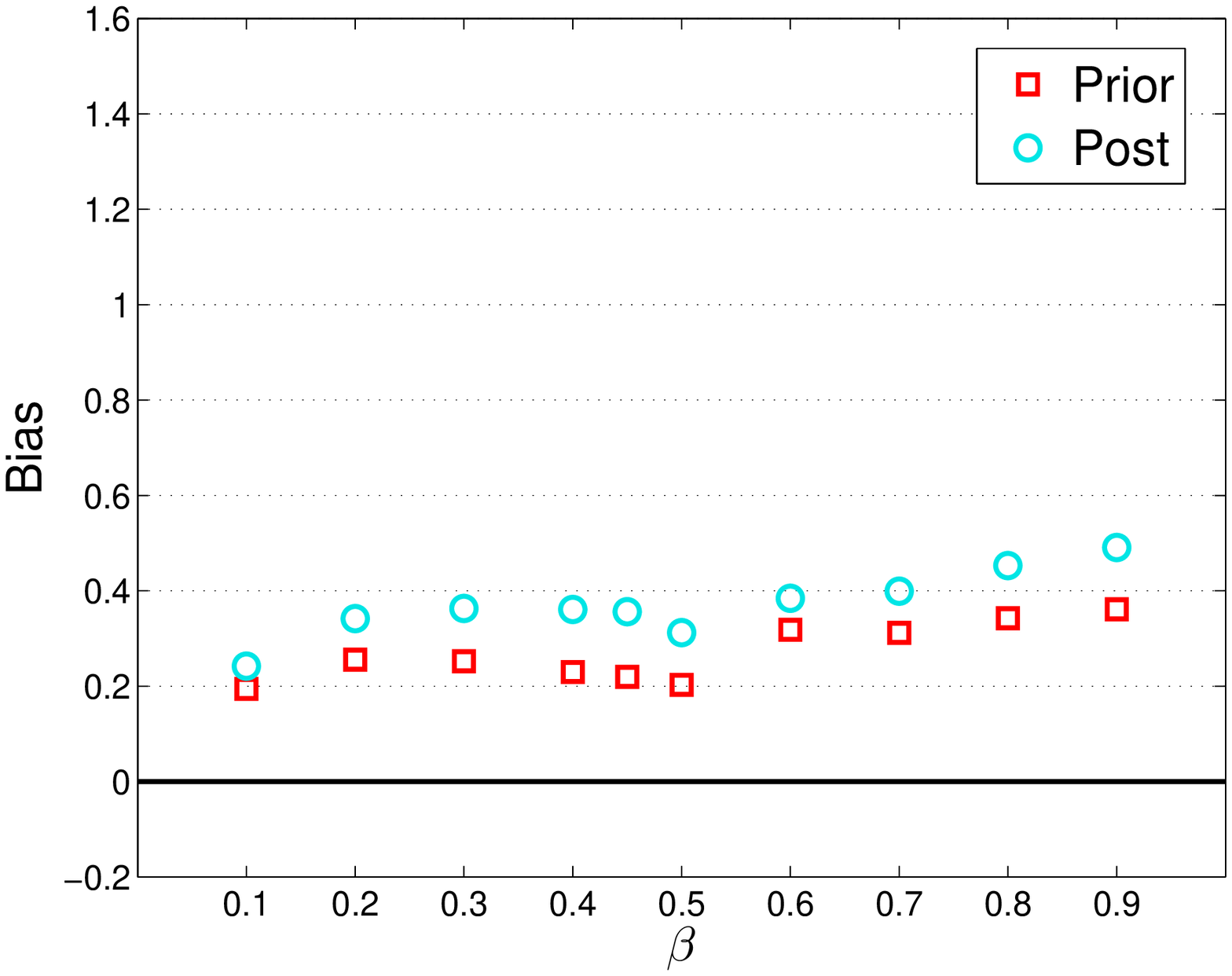}\\ \hline
&k=4&\\ \hline
\includegraphics[scale= 0.25]{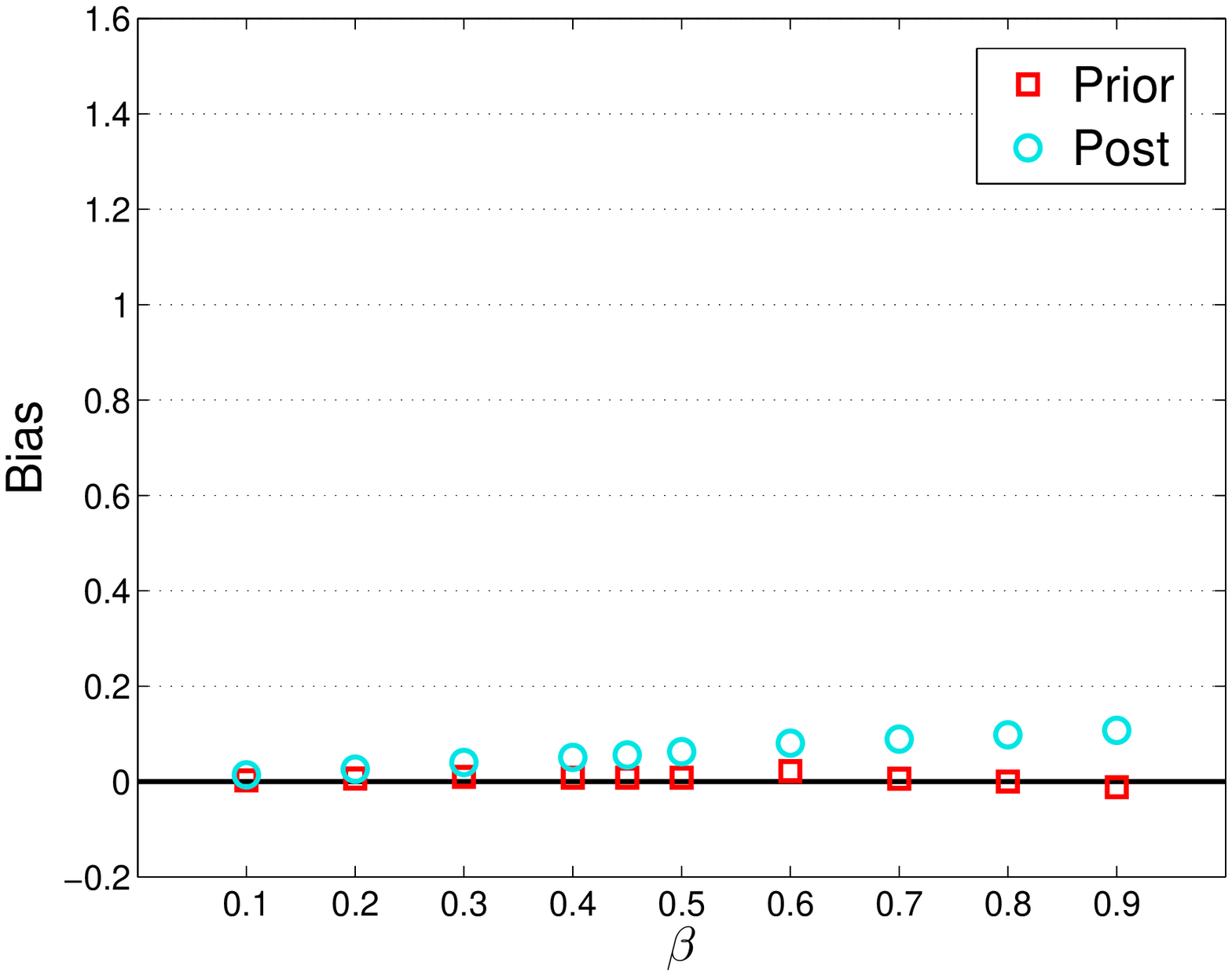}&\includegraphics[scale= 0.25]{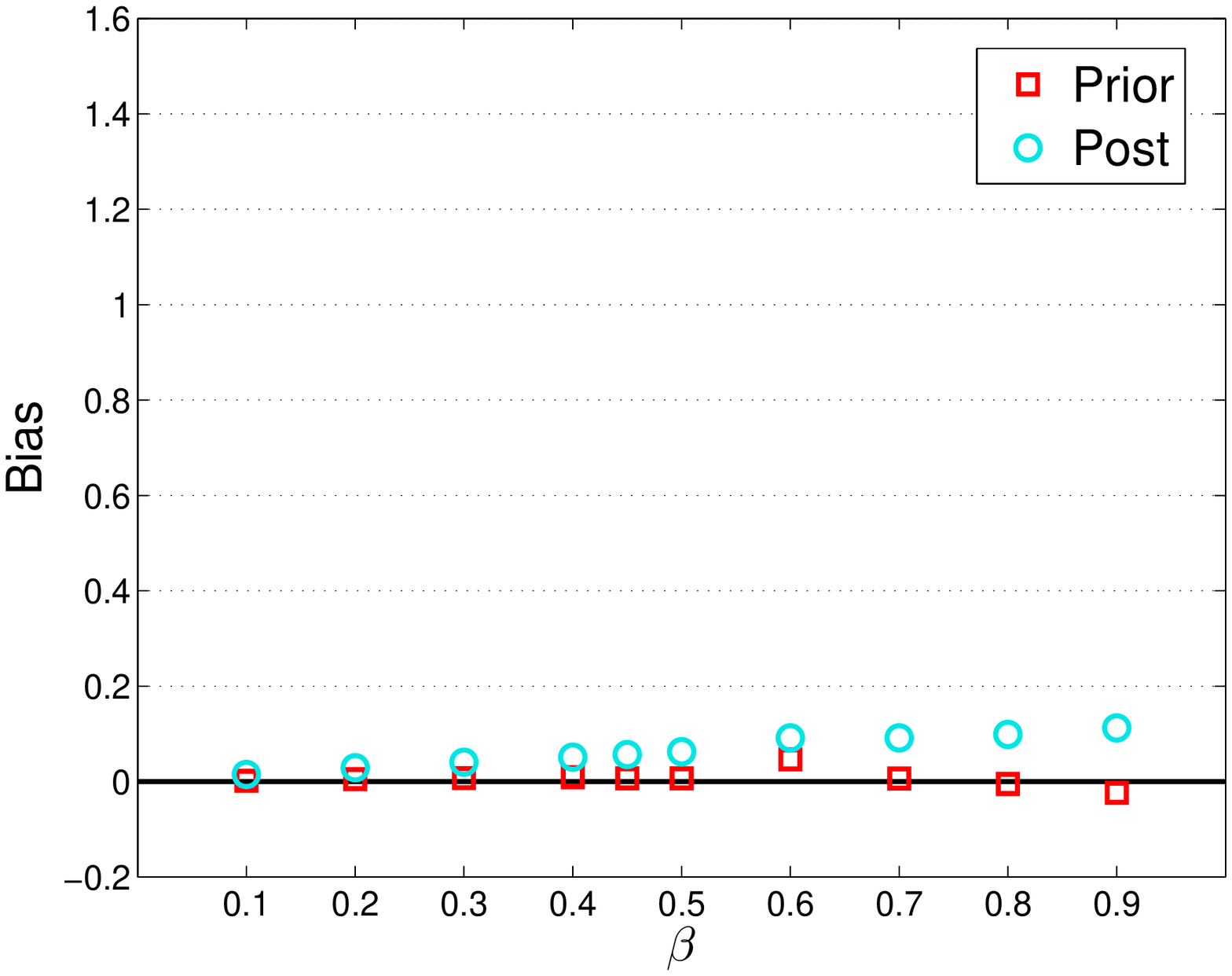}&
\includegraphics[scale= 0.25]{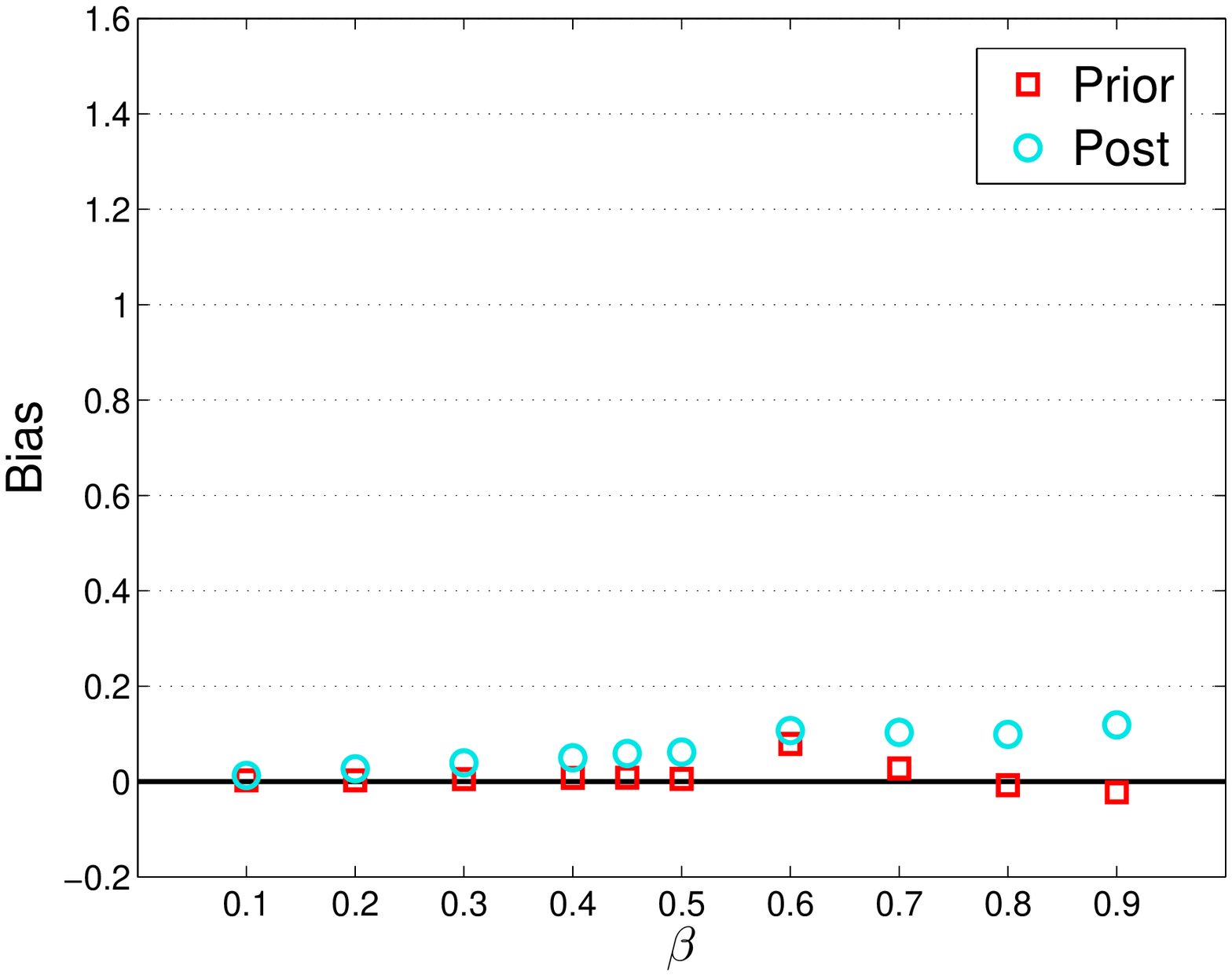}\\ \bottomrule
 \end{array} \]
\caption{Bias of the estimators produced by using $\widehat{\bm x}_{\mbox{\scriptsize ICM}}$ instead of the Potts model realization, for $L\in\{2,3,4\}$; $k\in\{1,4\}$; and several values of $\beta$.}
 \label{grafico_ICM_bias_vs_beta}
\end{figure*}

Fig.~\ref{grafico_ICM_bias_vs_beta} shows the bias of both estimators as a function of $\beta$, for different values of $L$ and $k$.
We should notice that both estimators have a positive bias.
Negative bias is only present for  large values of $\beta$.
Also, it is important to notice that $k$ has more influence than $L$ on the accuracy of the estimators.
When $k=4$, the number of classes has little to no influence in the bias of the estimators.
In all analyzed cases, the prior estimator has better accuracy than the posterior estimator.

\begin{figure}[htb]
\centering
\subfigure[$k=1$\label{fig_haces_ICM_1a}]{
  \includegraphics[scale = 0.24]{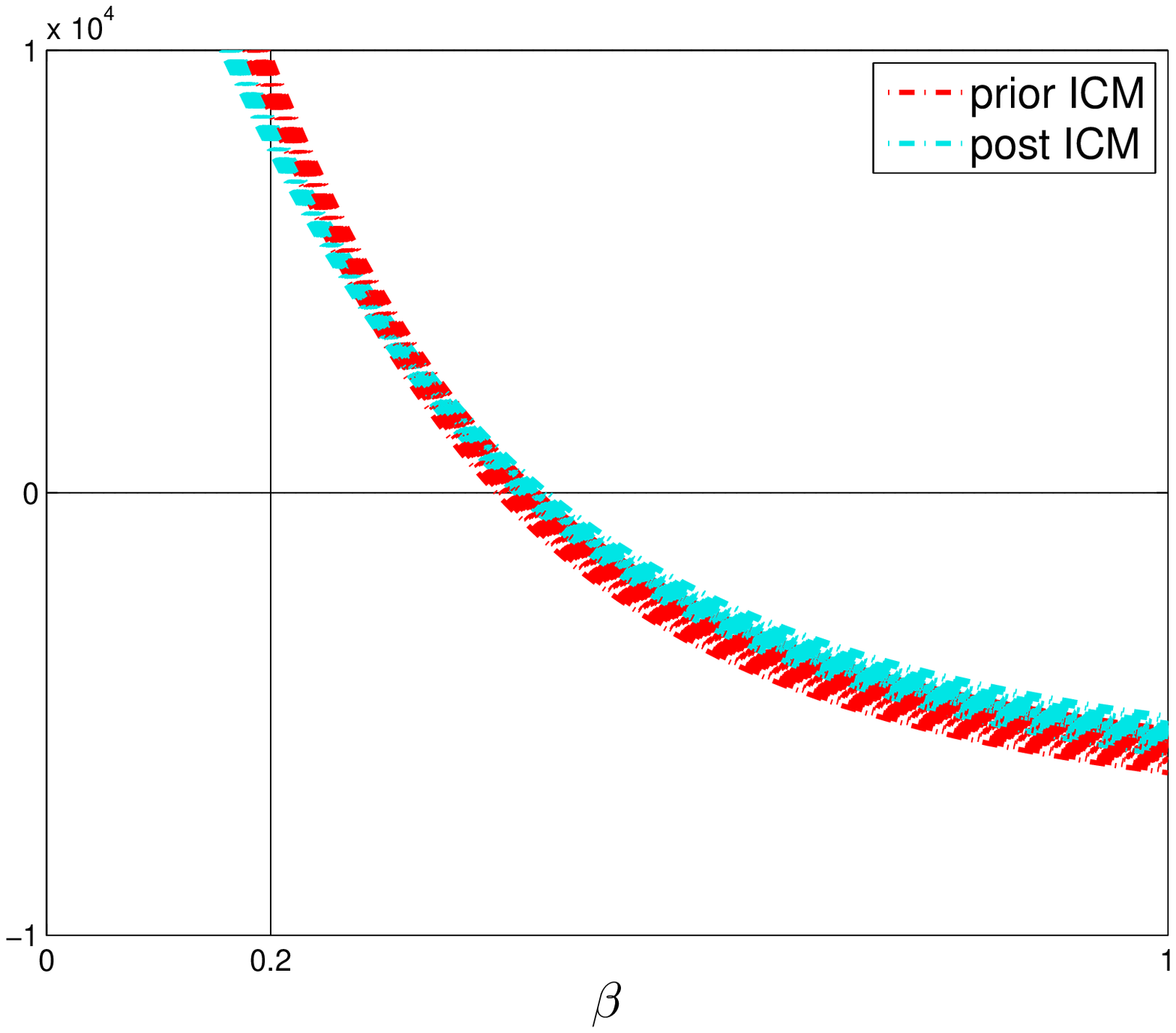}
}
\subfigure[$k=2$\label{fig_haces_ICM_1b}]{
  \includegraphics[scale = 0.24]{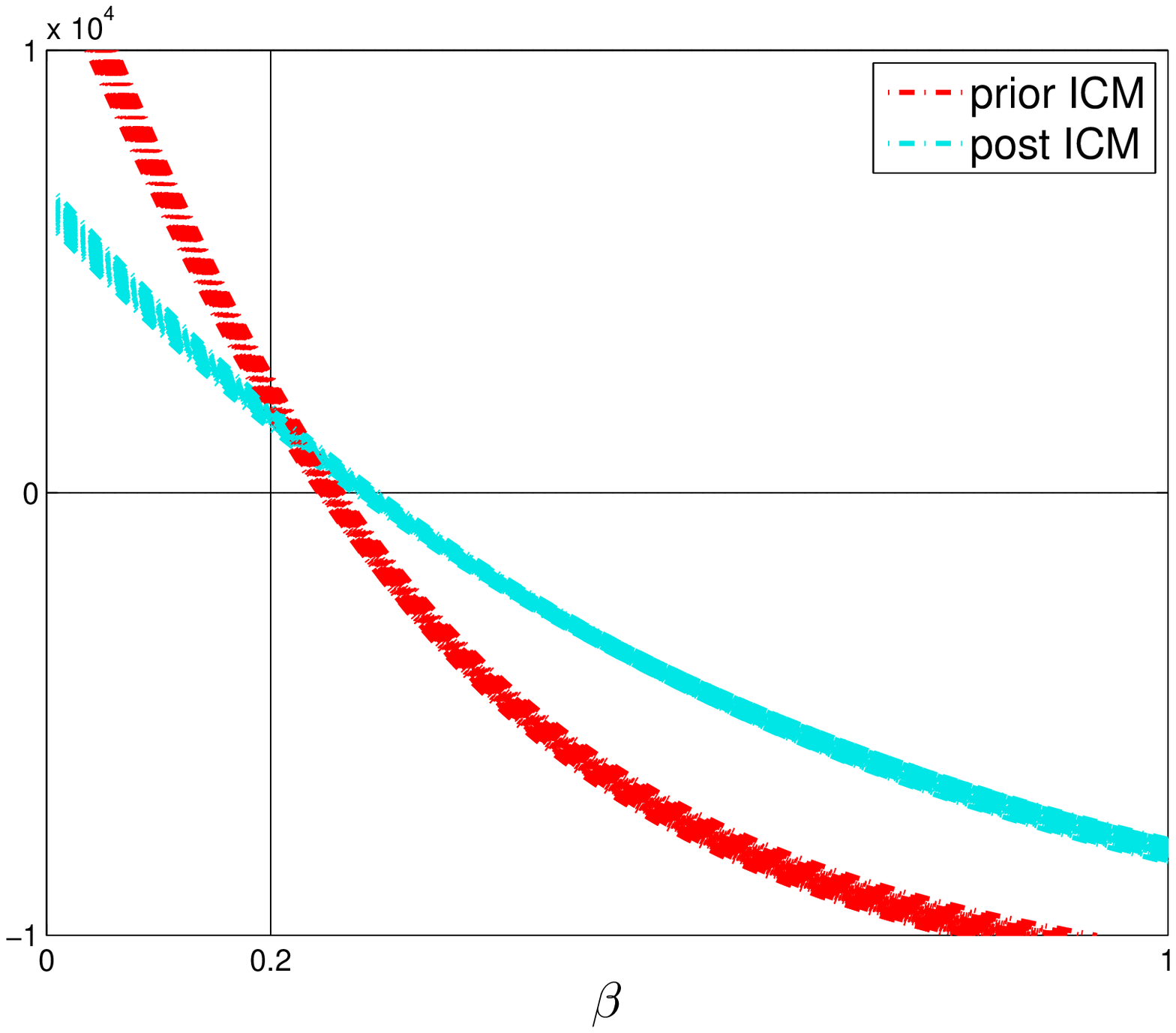}
}
\subfigure[$k=3$\label{fig_haces_ICM_1c}]{
  \includegraphics[scale = 0.24]{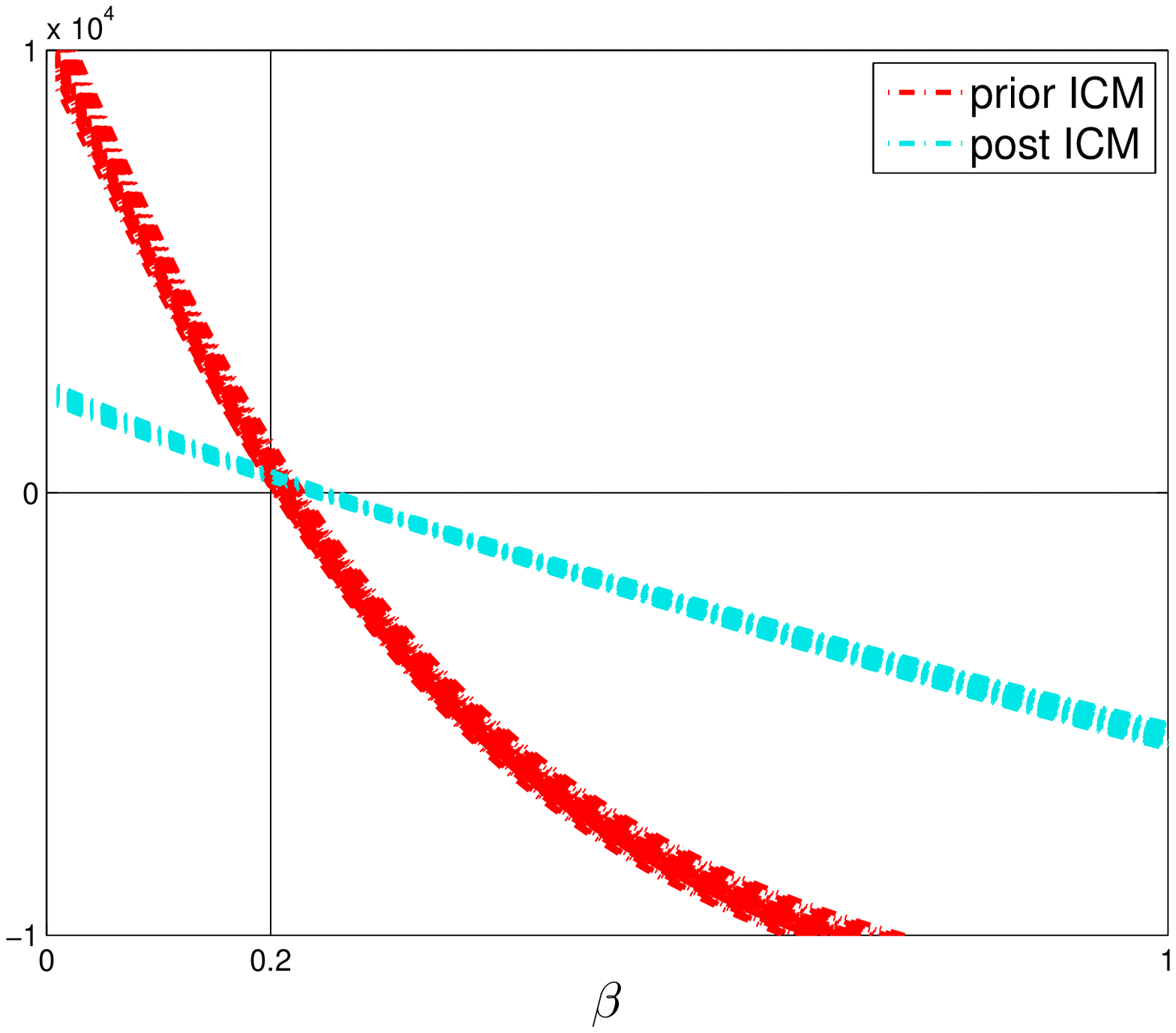}
}
\subfigure[$k=4$\label{fig_haces_ICM_1d}]{
  \includegraphics[scale = 0.24]{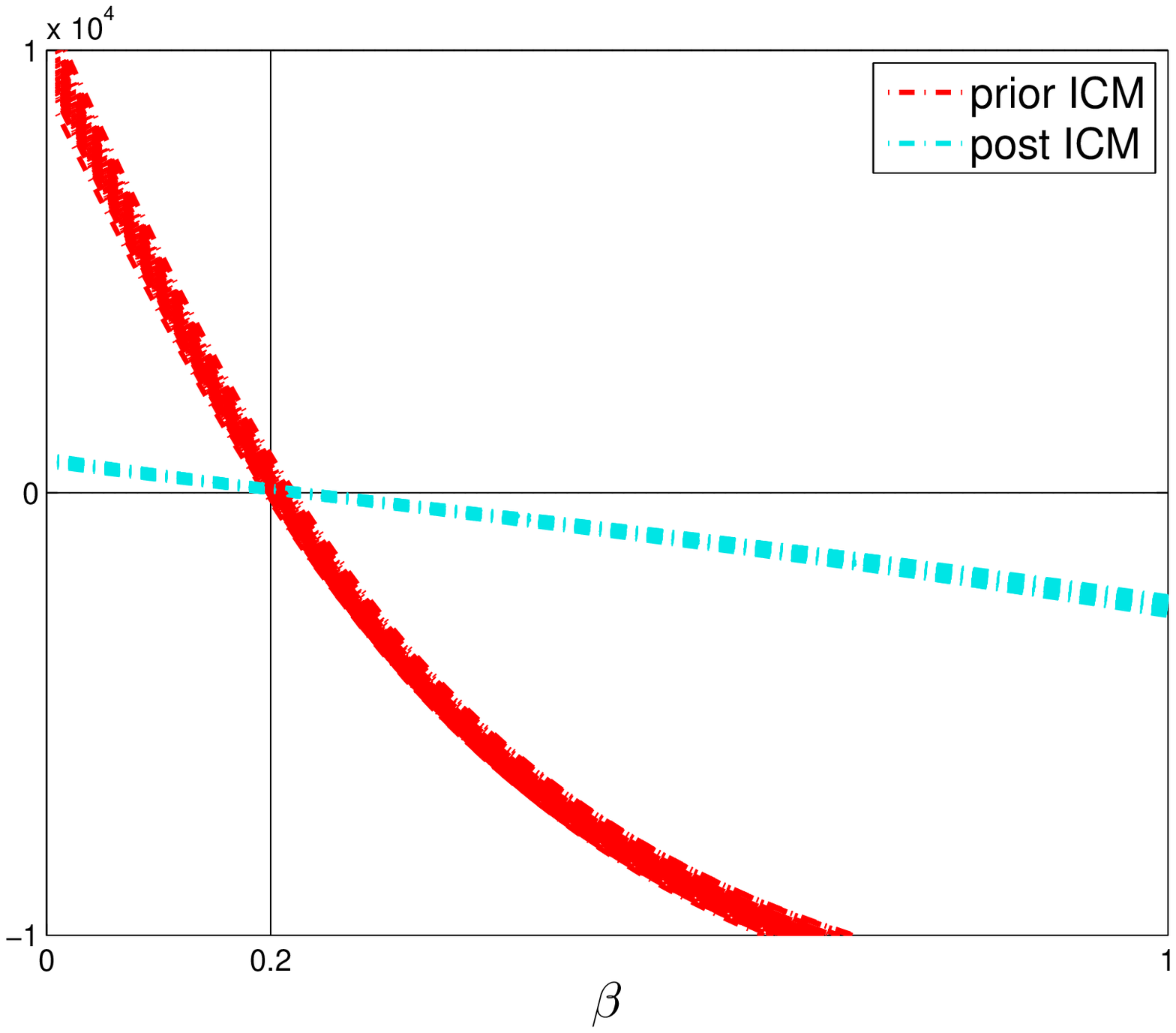}
}
\caption{Bundles of curves of the functions $f_{\mbox{\scriptsize prior}}^{\mbox{\scriptsize ICM}}$ and $f_{\mbox{\scriptsize post}}^{\mbox{\scriptsize ICM}}$ for $L=2$; $\beta=0.2$; and several values of $k$.}
 \label{grafico_haces_ICM}
\end{figure}

Such as we did in Fig.~\ref{fig_haces_fprior_fpost} and~\ref{grafico_haz_ML}, having $100$ replications of each case of number of classes, lags and $\beta$, we present curves that show the bias and variance of the estimators.
In Fig.~\ref{grafico_haces_ICM} we show the bundles of curves obtained for two classes,  $\beta=0.2$, for different lags.
We observe better estimation and reduced variance as the lag increases.



\section{Conclusion}\label{Sec:Conclusions}

We have analyzed the performance of two PML estimators of the smoothness parameter of the Potts model under simulation.
We report that under the true model, there is no statistical difference between the estimations.
But when we contaminated the model, introducing non contextual observations, or smoothed observations, the estimators showed differences in stability, bias and variance.

We have presented a theoretical analysis of such behavior, which leads us to conclude that, despite the reduced range of sampling of our simulation, our findings hold for other cases, allowing us to make the following statements regarding quality of PML estimation in the hidden Potts model case:
\begin{enumerate}
\item Radiometric unimodal distributions, regardless the number of true classes, produce severe bias in the PML estimators, which is reduced when ICM segmentation is considered as observed map.
\item Posterior PML estimators are roots of curves that flatten as functions of the difference of means in the radiometric information. This property combined with distortion produced by  dirty observed map of classes introduces a larger bias than prior PML estimators, which are not influenced by the radiometric distribution.
\item In the Hidden Potts Model problem, when there is no prior information about the smoothness of the map of classes, estimation should be made with the prior PML estimator, over ICM segmentation.
\end{enumerate}
The effect of adding the observed data into the estimation procedure does not improve the overall results. In fact, in some case the additional data worsens the estimation of the
smoothness parameter. This is more critical and noticeable in the bias, specially when the true value is relatively high.

Users of images with high signal-to-noise ratio should be specially cautious. As can be seen in Fig.~\ref{grafico_ML_bias_vs_beta} and Fig.~\ref{grafico_ICM_bias_vs_beta}, adding observed data increases the bias of the estimator of $\beta$,
and this effect is stronger the larger the value of $\beta$ is, i.e., the smoother the input map is.

If the use of additional data is not advisable in general for the estimation of the smoothness parameter, this is particularly important for users of relatively low resolution optical
data and with separable classes. These data yield (i) very smooth maps, and using the observed data would lead to highly biased estimates with negative results, and (ii) situations as
the one depicted for $L=4,k=4$ where the bias of the posterior estimator is much larger than the observed in the prior one.

We will prove elsewhere that posterior PML have interesting statistical properties such as consistency, asymptotic normality, as the prior estimators do.
Making use of extra information, without increasing complexity, they appear more intuitive than the PML estimators based only on prior information.
Nevertheless, inaccurate extra information produces unacceptable bias and distortion which should prevent the use of these new estimators in practical applications.

\section*{Appendix}

\subsection{Computational information}

Simulations were written on Matlab from scratch, and carried  on in a desktop computer with an Intel I5 2500 processor, and 8~GB of RAM memory.
A package with the routines is available for download from A.~G.~Flesia's Reproducible Research repository at Universidad Nacional de C\'ordoba.

\section*{Acknowledgments}
This work has been partially supported by  Argentinean grants ANPCyT-PICT 2008-00291 and Secyt UNC-PID 2012 05/B504.
J.~Gimenez was supported by a PhD student grant from Conicet.
A.~C.~Frery is grateful to CNPq and Fapeal.  Part of this work was included in the PhD dissertation of J.~Gimenez, under the direction of A.~G.~Flesia at the University of C\'ordoba,~Argentina~\cite{Gimenez2014t}.

\bibliographystyle{IEEEtran}

\begin{thebibliography}{10}
\providecommand{\url}[1]{#1}
\csname url@samestyle\endcsname
\providecommand{\newblock}{\relax}
\providecommand{\bibinfo}[2]{#2}
\providecommand{\BIBentrySTDinterwordspacing}{\spaceskip=0pt\relax}
\providecommand{\BIBentryALTinterwordstretchfactor}{4}
\providecommand{\BIBentryALTinterwordspacing}{\spaceskip=\fontdimen2\font plus
\BIBentryALTinterwordstretchfactor\fontdimen3\font minus
  \fontdimen4\font\relax}
\providecommand{\BIBforeignlanguage}[2]{{%
\expandafter\ifx\csname l@#1\endcsname\relax
\typeout{** WARNING: IEEEtran.bst: No hyphenation pattern has been}%
\typeout{** loaded for the language `#1'. Using the pattern for}%
\typeout{** the default language instead.}%
\else
\language=\csname l@#1\endcsname
\fi
#2}}
\providecommand{\BIBdecl}{\relax}
\BIBdecl
\bibitem{geman1984}
S.~Geman and D.~Geman, ``Stochastic relaxation, {G}ibbs distributions, and the
  {B}ayesian restoration of images,'' \emph{IEEE Trans. on Pattern Anal. and
  Mach. Intell.}, vol.~6, no.~6, pp. 721--741, Nov. 1984.

\bibitem{bustos1992}
O.~H. Bustos and A.~C. Frery, ``A contribution to the study of {M}arkovian
  degraded images: an extension of a theorem by {G}eman and {G}eman,''
  \emph{Comput. and Appl. Math.}, vol.~11, no.~3, pp. 281--285, Sept. 1992.

\bibitem{Ferrari1995}
P.~A. Ferrari, A.~Frigessi, and P.~G. de~S{\'a}, ``Fast approximate maximum a
  posteriori restoration of multicolor images,'' \emph{J. of the Roy. Stat.
  Soc.}, vol. B-57, no.~3, pp. 485--500, 1995.

\bibitem{besag1986}
J.~Besag, ``On the statistical analysis of dirty pictures,'' \emph{J. of the
  Roy. Stat. Soc.}, vol. B-48, no.~3, pp. 259--302, 1986.

\bibitem{Arbia1999}
G.~M. Arbia, R.~Benedetti, and G.~Espa, ``Contextual classification in image
  analysis: an assessment of accuracy of {ICM},'' \emph{Comput. Stat. \& Data
  Anal.}, vol.~30, no.~4, pp. 443--455, June 1999.

\bibitem{Jackson2002}
Q.~Jackson and D.~A. Landgrebe, ``Adaptive {B}ayesian contextual classification
  based on {M}arkov random fields,'' \emph{IEEE Trans. on Geosci. and Remote
  Sens.}, vol.~40, no.~11, pp. 2454--2463, Nov. 2002.

\bibitem{Descombes1999}
X.~Descombes, R.~D. Morris, J.~Zerubia, and M.~Berthod, ``Estimation of
  {M}arkov random field prior parameters using {M}arkov chain {M}onte {C}arlo
  maximum likelihood,'' \emph{IEEE Trans. on Image Process.}, vol.~8, no.~7,
  pp. 954--963, July 1999.

\bibitem{Melgani2003}
F.~Melgani and S.~B. Serpico, ``A {M}arkov random field approach to
  spatio-temporal contextual image classification.'' \emph{IEEE Trans. on
  Geosci. and Remote Sens.}, vol.~41, no.~11, pp. 2478--2487, Oct. 2003.

\bibitem{Tso1999}
B.~C. Tso and P.~M. Mather, ``Classification of multisource remote sensing
  imagery using a genetic algorithm and {M}arkov random fields.'' \emph{IEEE
  Trans. on Geosci. and Remote Sens.}, vol.~37, no.~3, pp. 1255--1260, May.
  1999.

\bibitem{Frery2007}
A.~C. Frery, A.~H. Correia, and C.~C. Freitas, ``Classifying multifrequency
  fully polarimetric imagery with multiple sources of statistical evidence and
  contextual information,'' \emph{IEEE Trans. on Geosci. and Remote Sens.},
  vol.~45, no.~10, pp. 3098--3109, Oct. 2007.

\bibitem{Frery2009}
A.~C. Frery, S.~Ferrero, and O.~H. Bustos, ``The influence of training errors,
  context and number of bands in the accuracy of image classification,''
  \emph{Int. J. of Remote Sens.}, vol.~30, no.~6, pp. 1425--1440, Mar. 2009.

\bibitem{McGrory2009}
C.~A. Mc{G}rory, D.~M. Titterington, R.~Reeves, and A.~N. Pettitt,
  ``Variational {B}ayes for estimating the parameters of a hidden {P}otts
  model,'' \emph{Stat. and Comput.}, vol.~19, no.~3, pp. 329--340, Sept. 2009.

\bibitem{liu2000}
J.~Liu, L.~Wang, and S.~Li, ``{MRF} parameter estimation by {MCMC} method,''
  \emph{Pattern Recognition}, vol.~33, no.~11, pp. 1919--1925, Nov. 2000.

\bibitem{Risser2011}
L.~Risser, T.~Vincent, F.~Forbes, J.~Idier, and P.~Ciuciu, ``Min-max
  extrapolation scheme for fast estimation of {3D} {P}otts field partition
  functions. {A}pplication to the joint detection-estimation of brain activity
  in f{MRI},'' \emph{J. of Signal Process. Syst.}, vol.~65, no.~3, pp.
  325--338, Dec. 2011.

\bibitem{Ibanez2003}
M.~V. Ib\'{a}\~{n}ez and A.~Sim\'{o}, ``Parameter estimation in {M}arkov random
  field image modeling with imperfect observations: a comparative study,''
  \emph{Pattern Recognition Lett.}, vol.~24, no.~14, pp. 2377--2389, Oct. 2003.

\bibitem{Ali2008}
A.~M. Ali, A.~A. Farag, and G.~L. {Gimel Farb}, ``Analytical method for {MGRF}
  {P}otts model parameter estimation,'' \emph{In proc. of: 19th Int. Conf. on
  Pattern Recognition (ICPR 2008)}, pp. 1--4, Dec. 2008.


\bibitem{Pereyra2013}
M.~Pereyra, N.~Dobigeon, H.~Batatia, and J.~Tourneret, ``Estimating the
  granularity parameter of a {P}otts-{M}arkov random field within an {MCMC}
  algorithm,'' \emph{IEEE Trans. on Image Process.}, vol.~22, no.~6, pp.
  2385--2397, June 2013.

\bibitem{Besag1975}
J.~Besag, ``Statistical analysis of non-lattice data,'' \emph{J. of the Roy.
  Stat. Soc. Series D (The Statistician)}, vol.~24, no.~3, pp. 179--195, Sept.
  1975.

\bibitem{Levada2008}
A.~L.~M. Levada, N.~D.~A. Mascarenhas, and A.~Tann{\'u}s, ``Pseudolikelihood
  equations for the {P}otts {MRF} model parameters estimation on higher order
  neighborhood systems,'' \emph{IEEE Geosci. and Remote Sens. Lett.}, vol.~5,
  no.~3, pp. 522--526, July 2008.

\bibitem{Flesia2013a}
 A.~G. Flesia,  J.~Gimenez and  J.~Baumgartner, ``On segmentation with {M}arkovian models,''
 \emph{In proc. of: XIV Argentine Symposium on Artificial Intelligence (ASAI 2013)}, Sept. 2013.

\bibitem{Flesia2013b}
 A.~G. Flesia,  J.~Baumgartner, J.~Gimenez and J.~Martinez ``Accuracy of {MAP} segmentation with hidden Potts and Markov mesh prior models via Path Constrained Viterbi Training, Iterated Conditional Modes and Graph Cut based algorithms,''
 \emph{arXiv preprint arXiv:1307.2971 (2013)}.

\bibitem{Gimenez2013}
J.~Gimenez, A.~C. Frery, and A.~G. Flesia, ``Inference strategies for the
  smoothness parameter in the {P}otts {M}odel,'' \emph{In proc. of: Geoscience and Remote Sensing Symposium (IGARSS)}, pp. 2539--2542, July 2013.

\bibitem{Levada2009}
A.~L.~M. Levada, N.~D.~A. Mascarenhas, and A.~Tann\'us, ``Pseudo-likelihood
  equations for {P}otts model on higher-order neighborhood systems: A
  quantitative approach for parameter estimation in image analysis,''
  \emph{Brazilian J. of Prob. and Stat.}, vol.~23, no.~2, pp. 120--140, 2009.

\bibitem{SW1987}
R.~H. Swendsen and J.~S. Wang, ``Nonuniversal critical dynamics in {M}onte
  {C}arlo simulations,'' \emph{Physical Review Lett.}, vol.~58, no.~2, pp.
  86--88, Jan 1987.

\bibitem{Gimenez2014t}
\BIBentryALTinterwordspacing
J.~Gimenez, ``Estimaci\'on de par\'ametros de modelos a priori para
  segmentaci\'on contextual de im\'agenes,'' Ph.D. dissertation, Univ. Nac. de
  C\'ordoba, C\'ordoba, Argentina, 2014. [Online]. Available:
  \url{http://www.famaf.unc.edu.ar/wp-content/uploads/2014/04/DMat84.pdf}
\BIBentrySTDinterwordspacing

\end{thebibliography}

\end{document}